\documentclass[12pt]{article}
\usepackage{epsfig, amsmath, amssymb}
\usepackage{graphicx,psfrag}
\usepackage{xcolor} 
\newcommand{\nn}{\nonumber}
\newcommand{\be}{\begin{equation}}
\newcommand{\ee}{\end{equation}}
\newcommand{\ben}{\begin{equation}}
\newcommand{\een}{\end{equation}}
\newcommand{\bea}{\begin{eqnarray}}
\newcommand{\eea}{\end{eqnarray}}
\newcommand{\bA}{\begin{array}}
\newcommand{\eA}{\end{array}}
\newcommand{\bc}{\begin{center}}
\newcommand{\ec}{\end{center}}
\newcommand{\al}{\alpha}

\newcommand{\ra}{\rightarrow}
\newcommand{\del}{\partial}

\newcommand{\ie}{{\it i.e.}}
\newcommand{\eg}{{\it e.g.}}

\def\BZ{{\mathbb Z}}

\newcommand{\lan}{\langle}
\newcommand{\ran}{\rangle}

\begin{document}


\begin{titlepage}

\bc

\hfill 

\vspace{25mm}


{\Huge Cosmological singularities, entanglement 
  \\ [2mm]
  and quantum extremal surfaces} 
\vspace{16mm}

{\large A. Manu,\ \ K.~Narayan,\ \ Partha Paul} \\
\vspace{3mm}
{\small \it Chennai Mathematical Institute, \\}
{\small \it SIPCOT IT Park, Siruseri 603103, India.\\}

\ec
\vspace{35mm}

\begin{abstract}
We study aspects of entanglement and extremal surfaces in various
families of spacetimes exhibiting cosmological, Big-Crunch,
singularities, in particular isotropic $AdS$ Kasner. The classical
extremal surface dips into the bulk radial and time
directions. Explicitly analysing the extremization equations in the
semiclassical region far from the singularity, we find the surface
bends in the direction away from the singularity. In the 2-dim
cosmologies obtained by dimensional reduction of these and other
singularities, we have studied quantum extremal surfaces by
extremizing the generalized entropy. The resulting extremization shows
the quantum extremal surfaces to always be driven to the semiclassical
region far from the singularity. We give some comments and
speculations on our analysis.
\end{abstract}

\end{titlepage}

{\tiny 
\begin{tableofcontents}
\end{tableofcontents}
}


\section{Introduction}

Some very exciting discoveries have been made recently on the black
hole information paradox
\cite{Penington:2019npb,Almheiri:2019psf,Almheiri:2019hni}, unravelled
via the study of entanglement and quantum extremal surfaces: a
qualitative review is \cite{Almheiri:2020cfm}.  Perhaps the central
point is that the generalized entropy
\cite{Faulkner:2013ana,Engelhardt:2014gca} obtained by incorporating
the bulk entanglement entropy of matter to the classical area of the
entangling RT/HRT surface \cite{Ryu:2006bv}-\cite{Rangamani:2016dms}
makes a qualitative difference to the location of the quantum extremal
surfaces, with explicit calculation possible in effective
2-dimensional models where the bulk entanglement entropy can be
studied through 2-dim CFT techniques (see also the early work
\cite{Fiola:1994ir}). A noteworthy point in these studies is that
apparently no information on the singularity inside the black hole and
associated stringy/quantum gravity effects is necessary: this is
perhaps not surprising since the near horizon region is adequately
semiclassical but is striking. The interior singularity could be
regarded as a cosmological, spacelike Big-Crunch singularity.  In this
light, or indeed independently, it is tempting to ask if quantum
extremal surfaces might be used to probe cosmological, Big-Crunch or
-Bang, singularities: a priori it is not clear if this makes sense
since the near singularity region is expected to be rife with severe
stringy/quantum gravity effects. But what one might hope is to gain
some insight into how these extremal surfaces either probe or avoid
such singularities, in the process learning more about entanglement
and quantum extremal surfaces in general. Some interesting recent work
on quantum extremal surfaces and cosmologies appears in
\cite{Chen:2020tes,Hartman:2020khs} and also
\eg\ \cite{Krishnan:2020fer,VanRaamsdonk:2020tlr,Balasubramanian:2020xqf,Sybesma:2020fxg}.

In this paper we investigate aspects of entanglement and quantum
extremal surfaces in certain classes of spacetimes exhibiting
cosmological singularities studied first in
\cite{Das:2006dz}-\cite{Awad:2008jf}. These are time-dependent
deformations of $AdS/CFT$
\cite{Maldacena:1997re}-\cite{Aharony:1999ti} where the bulk develops
spacelike Big-Crunch singularities when the dual field theory is
deformed to be on a time-dependent space alongwith a time-dependent
gauge coupling. Perhaps the simplest of these is the $AdS$ Kasner
spacetime.  There are no horizons in these spacetimes.  In
\cite{Bhattacharya:2020qil}, certain families of 2-dim cosmologies
with cosmological singularities were studied in 2-dim dilaton gravity
with an extra scalar which drives the dynamics in these theories.
Some of these can be thought of as the dimensional reduction of the
isotropic $AdS$ Kasner and other, more general, cosmologies with
Big-Crunch singularities.

First we study aspects of classical extremal RT/HRT surfaces in the
higher dimensional backgrounds: due to the time dependence the
surfaces also dip in the time direction (besides the holographic
radial direction). By a detailed study of the extremization equations
in the reliable semiclassical regime far from the singularity, we show
that the classical entangling surface has only mild time dependence
and bends away from the singularity. Next we study quantum extremal
surfaces in the 2-dim cosmologies mentioned above, keeping in mind the
key quantitative feature that the 2-dim backgrounds here allow
formulating the bulk entanglement contribution in terms of well-known
2-dim CFT techniques.  Assuming that the bulk matter in the region far
from the singularity is approximately in the ground state is
reasonable since the time variations are small there: then the bulk
entanglement can be approximated using these techniques. The
time-dependence inherent in these backgrounds then leads automatically
to an extrapolation to the the rest of the spacetime. Extremizing the
resulting generalized entropy then shows that the quantum extremal
surfaces are also driven to the semiclassical region far from the
singularity. We discuss various features here and implications.

In sec.~2, we review certain aspects of these cosmological
singularities and in particular the 2-dim ones in
\cite{Bhattacharya:2020qil}. In sec.~3, we discuss aspects of RT/HRT
surfaces in the higher dimensional cosmologies, with primary focus on
the $AdS$ Kasner case. In sec.~4, we discuss quantum extremal
surfaces: after various generalities, we discuss some time-independent
backgrounds which provide some intuition (sec.~4.1), and then 2-dim
cosmologies with the $AdS$ Kasner singularity in sec.~4.2 and more
general ones in sec.~4.3. We close with a Discussion in sec.~5, and
some technical details in Appendices.

\section{Cosmological singularities}\label{sec:rev2dCos}

Here we review some aspects of the cosmological spacetimes discussed
in \cite{Bhattacharya:2020qil}. The higher dimensional backgrounds
were studied long back as time-dependent deformations of $AdS/CFT$ in
\cite{Das:2006dz}-\cite{Awad:2008jf} towards gaining
insights via gauge/gravity duality into cosmological (Big-Bang or
-Crunch) singularities: further investigations on some of these appear
in \eg\ \cite{Engelhardt:2014mea}-\cite{Engelhardt:2016kqb}:\ some
reviews of cosmological singularities in string theory appear in
\eg\ \cite{Craps:2006yb,Burgess:2011fa}. While the bulk spacetime
develops a cosmological Big-Crunch (or -Bang) singularity and breaks
down, the holographic dual field theory (in the $AdS_5$ case), living
on a space that itself crunches, is subject to a severe time-dependent
gauge coupling $g_{YM}^2=e^\Psi$ and may be hoped to provide insight
into the dual dynamics. In this case the scalar $\Psi$ controls the
gauge/string coupling. Generically it was found that the gauge theory
response also ends up being singular \cite{Awad:2008jf}. There is a
large family of such backgrounds exhibiting cosmological
singularities, some of which we will review below. Various other
references are listed in \cite{Bhattacharya:2020qil}.

Some of these backgrounds have the technical feature that the spatial
directions are all on the same footing: this isotropy allows studying
these backgrounds from a possibly simpler perspective.  In
\cite{Bhattacharya:2020qil} a dimensional reduction on the spatial
part of these backgrounds was carried out, which enables recasting
these backgrounds from the point of view of 2-dim dilaton gravity 
with a dilaton potential and an extra scalar that drives the dynamics
in a nontrivial manner.
A prototypical example of this is the $AdS_D$ Kasner spacetime
\cite{Das:2006dz} and the 2-dim cosmology obtained from its reduction
\cite{Bhattacharya:2020qil}: see (\ref{AdSDK-2d}).
More generally the higher dimensional space and its reduction
ansatz are of the form
\be\label{redux+Weyl}
ds^2_D = g^{(2)}_{\mu\nu} dx^\mu dx^\nu + \phi^{2\over d_i} d\sigma_{d_i}^2\ ;
\qquad\quad g_{\mu\nu}=\phi^{{d_i-1\over d_i}} g^{(2)}_{\mu\nu}\ ,
\qquad D=d_i+2\ .
\ee
The Weyl transformation from $g^{(2)}_{\mu\nu}$ to the 2-dim metric
$g_{\mu\nu}$ ensures that the dilaton kinetic energy vanishes and
the action becomes
\be\label{actionXPsiU}
S= {1\over 16\pi G_2} \int d^2x\sqrt{-g}\, \Big(\phi\mathcal{R}
- U(\phi,\Psi) -\frac{1}{2} \phi (\partial\Psi)^2 \Big)\ ,
\ee
The dilaton potential $U(\phi,\Psi)$ now possibly couples the dilaton
$\phi$ to $\Psi$.\ Certain aspects of generic dilaton gravity theories
of this kind (and these 2-dim cosmological backgrounds), dimensional
reduction and holography are discussed in \cite{Narayan:2020pyj}. See
\eg\ \cite{Strominger:1994tn} for early discussions of 2-dim dilaton
gravity in the context of 2-dim black holes as well as
\cite{Almheiri:2014cka} in the context of $AdS_2$ holography. We obtain
\bea\label{2dimseom-EMD0-Psi}
g_{\mu\nu}\nabla^2\phi-\nabla_{\mu}\nabla_{\nu}\phi
  +\frac{g_{\mu\nu}}{2}\Big(\frac{\phi}{2}(\partial\Psi)^2+U\Big)
  -\frac{\phi}{2}\partial_{\mu}\Psi\partial_{\nu}\Psi = 0\ ,\qquad\quad &&
  \nonumber\\ [1mm]
\mathcal{R}-\frac{\partial U}{\partial\phi}
  -\frac{1}{2}(\partial\Psi)^2 = 0\ ,\qquad\qquad
\frac{1}{\sqrt{-g}}\partial_{\mu}(\sqrt{-g}\,\phi \partial^{\mu}\Psi)
  -\frac{\partial U}{\partial\Psi} = 0\ ,&&
\eea
as the equations of motion. These give in conformal gauge
$g_{\mu\nu}=e^f\eta_{\mu\nu}$\,:
\bea\label{2dimseom-EMD1-Psi}
(tr)&& \qquad  \del_t\del_r\phi 
- {1\over 2} f'\del_t\phi - {1\over 2} {\dot f} \del_r\phi
+ {\phi\over 2}{\dot\Psi} \Psi' = 0\ , \nn\\
(rr+tt) && \ \ \
- \del_t^2\phi - \del_r^2\phi + {\dot f} \del_t\phi + f' \del_r\phi
- {\phi\over 2} ({\dot\Psi})^2 - {\phi\over 2} (\Psi')^2 = 0 ,\qquad \nn\\
[1mm]
(rr-tt) && \ \ \
- \del_t^2\phi + \del_r^2\phi + e^f U = 0\ ,\\ [1mm]
(\phi)&& \quad\  \big( {\ddot f} - f'' \big)
- {1\over 2} (-({\dot\Psi})^2+(\Psi')^2)
- e^f \frac{\partial U}{\partial\phi} = 0 ,\nn \\
(\Psi)&& \quad\   - \del_{t}(\phi\del_t\Psi) + \del_{r}(\phi\del_r\Psi)
- e^f {\del U\over\del\Psi} = 0\ . \nn 
\eea
There is nontrivial dynamics in the theory (\ref{actionXPsiU}) driven
by the extra scalar $\Psi$. In particular there are nontrivial
cosmological singularity solutions here, which were analysed in
\cite{Bhattacharya:2020qil}. The power-law scaling ansatze for the
2-dim fields and the higher dimensional spacetimes, from which these
can be thought of as arising from via reduction, are
\be\label{phie^fPsi-ansatz}
\phi=t^kr^m,\quad e^f=t^ar^b,\quad e^\Psi=t^\al r^\beta \quad\ra\quad
ds_D^2 = {e^f\over \phi^{(d_i-1)/d_i}}\big(-dt^2+dr^2\big)+\phi^{2/d_i}dx_i^2\ .
\ee
Note that $r=0$ is the asymptotic (holographic) boundary.
In the vicinity of the Big-Crunch singularity, there is rapid time
variation, approaching a divergence.
Thus taking the time derivative terms to be dominant (dropping all the
other terms) gives the near singularity behaviour described by
\be\label{nearSing-EMD-1}
- \del_t^2\phi + {\dot f} \del_t\phi - {\phi\over 2} ({\dot\Psi})^2 \sim 0 ,
\qquad  - \del_t^2\phi \sim 0 ,\qquad 
 {\ddot f} + {1\over 2} ({\dot\Psi})^2 \sim 0 ,\qquad
- \del_{t}( \phi \del_t\Psi) \sim 0 .\ \ 
\ee
This appears ``universal'': the dilaton potential $U$ governing the
asymptotic behaviour of the background has disappeared.
Solving these shows a ``universal'' subsector
\be\label{univSing}
\phi \sim t ,\quad\ e^f\sim t^a ,\quad\ e^\Psi\sim t^\al ;\qquad 
a={\al^2\over 2}\ ,
\ee
which governs the cosmological singularity.
Using (\ref{phie^fPsi-ansatz}), various families of nontrivial 2-dim
cosmologies can be found as exact classical solutions: in the vicinity
of the singularity they vindicate this universal behaviour but far
from this region exhibit various kinds of asymptotic data which is
encoded by the dilaton potential $U$. Some noteworthy examples are:
\begin{itemize}
\item{Flat space: $U=0$. We obtain
\be\label{flat-2d}
\phi=t ,\quad ds^2=t^{\al^2/2} (-dt^2+dr^2) ,\quad e^\Psi=t^\al\ ,
\ee
With $t=T^{1-p_1}$, these can be seen to be the reduction of ``mostly
isotropic'' Kasner singularities\
$ds^2=-dt^2+t^{2p_1}dx_1^2+t^{2p_2}\sum_idx_i^2$.}
\item{$AdS$ Kasner spacetimes: these are of the form 
\bea\label{AdSDK-2d}
&& U=2\Lambda\phi^{1/d_i}\,,\quad \Lambda=-{1\over 2}\,d_i(d_i+1)\ ,\qquad
p={1\over d_i}\ , \quad
\al = \sqrt{{2(d_i-1)\over d_i}} , \nn\\
&&  ds^2 = {R^2\over r^2} (-dt^2 + dr^2) + {t^{2p}\,R^2\over r^2} dx_i^2\,,
\qquad e^\Psi = t^\al\,,\qquad d_ip^2=1-{1\over 2}\al^2\ , \nn\\ [1mm]
\ra\ \ && \phi={t\,R^{d_i}\over r^{d_i}}\,,\qquad
ds^2={t^{(d_i-1)/d_i}\,R^{d_i+1}\over r^{d_i+1}}(-dt^2+dr^2)\,,\qquad
  e^\Psi=t^{\sqrt{2(d_i-1)/d_i}}\ .\quad
\eea
$R$ is the $AdS$ length scale. We are suppressing an implicit Kasner
scale $t_K$: \eg\ $t^{2p}\ra (t/t_K)^{2p}$. We will reinstate this as
required. The higher dimensional spacetimes and their dual field
theories were studied in \cite{Das:2006dz}-\cite{Awad:2008jf}.
}
\item{Hyperscaling violating cosmologies: the 2-dim and higher
  dimensional backgrounds are of the form (\ref{phie^fPsi-ansatz}) with
  exponents and parameters:
\bea\label{hvL-2d}
&& U(\phi,\Psi)=2\Lambda\phi^{{1\over d_i}} e^{\gamma\Psi}\,,\quad
\Lambda = -{1\over 2}(d_i+1-\theta)(d_i-\theta) ,\quad
\gamma = {-2\theta\over\sqrt{2d_i(d_i-\theta)(-\theta)}}\ , \nn\\
&& \qquad
m=-(d_i-\theta)\ ,\quad b={m(1+d_i)\over d_i}\ ,\quad \beta=-m\gamma\,,\nn\\
&& \qquad k=1 ,\quad a={\al^2\over 2}\,,\quad
\al=-\gamma \pm \sqrt{\gamma^2+{2(d_i-1)\over d_i}}\ .
\eea
Here $\theta<0$, $\gamma>0$. The higher dimensional backgrounds here
can be obtained as cosmological deformations of reductions of
nonconformal branes down to $D$ dimensions.\\
There are also still more complicated hyperscaling violating Lifshitz
cosmologies (with nontrivial Lifshitz exponents $z$ as well) and their
reductions down to 2-dimensions which were obtained in
\cite{Bhattacharya:2020qil}: we will not discuss them here.
}
\end{itemize}
In what follows we will study these cosmological backgrounds and the
2-dim cosmologies obtained from their reduction. In certain places we
will find it convenient and instructive to focus on the $AdS$ Kasner
singularities (\ref{AdSDK-2d}) above and the corresponding 2-dim
cosmologies. In the following it will be useful to note the general
cosmological solutions in the form (\ref{phie^fPsi-ansatz}), with the
2-dim fields on the left, and the higher dimensional spacetime on the
right.

\section{Classical extremal surfaces}

We would like to study the behaviour of RT/HRT surfaces in these
cosmological backgrounds. The nontrivial time-dependence here in
general complicates finding closed form expressions for the extremal
surfaces but nevertheless various scaling results and intuition can be
obtained from this study, as well as quantitative information in the
semiclassical regime far from the singularity.

In the higher dim isotropic spacetime (\ref{phie^fPsi-ansatz}), since
all $x_i$ are equivalent, let us consider a strip-shaped subsystem with
width $l$ along the $x\in x_i$ direction and wrapping all other $x_i$
directions.  The bulk extremal surface is anchored at its boundary and
dips into the bulk radial direction but also in time, essentially
forced by the time-dependence of the bulk cosmology. The surface
will continue to be spacelike: this is a nontrivial statement in such
time-dependent backgrounds.
The surface parametrisation and boundary conditions are
\be\label{RTsurf-bdycndns}
         {\cal S} \equiv (t(r), x(r))\ ;\qquad \Delta x = l\ ;\qquad
         t(r)\xrightarrow{r\ra 0} t_0\ .
\ee
For simplicity, we will consider the surface to be anchored on a $t=const$
slice on the boundary. The bulk surface can dip nontrivially
in time so $t(r)$ is potentially a nontrivial function starting at $t=t_0$
on the boundary, dipping into the bulk till some point then turning back
to the boundary again at $t_0$ (as we will see). More generally one
could consider ``tilted'' subsystems (with $t_L(0)\neq t_R(0)$): we
will not consider this. The area functional is\ \ (note $j$ takes
$D-3=d_i-1$ values)
\bea\label{EEtxr-areaFnal}
S &=& {1\over 4G_{d_i+2}} \int \prod^{j\in (1\ldots d_i-1)}_{x_j\neq x}
\left( \phi^{1/d_i} dx_j\right)
\sqrt{{e^f\over \phi^{(d_i-1)/d_i}}\big(-dt^2+dr^2\big)+\phi^{2/d_i}dx^2} \nn\\
&=&  {V_{d_i-1}\over 4G_{d_i+2}} \int dr\, \phi\,
\sqrt{{e^f\over \phi^{(d_i+1)/d_i}}\big(1-(\partial_{r}t)^{2}\big)
  +(\partial_{r}x)^{2}} \ \ .
\eea
There is no $x$-dependence so the momentum conjugate to $\del_rx$ gives
a conserved quantity  
\be
{\phi\, \del_r x\over\sqrt{{e^f\over \phi^{(d_i+1)/d_i}}\big(1-(\partial_{r}t)^{2}\big)+(\partial_{r}x)^{2}}} = const = A\ \
\ee
This gives
\be\label{extSurf-Xf}
(\del_rx)^2 = A^2\,{{e^f\over \phi^{(d_i+1)/d_i}}\big(1-(\partial_{r}t)^{2}\big)
  \over \phi^2-A^2}\ ,\qquad
S = {V_{d_i-1}\over 4G_{d_i+2}} \int dr\; 
{e^{f/2}\,\phi^{(3-1/d_i)/2}\over \sqrt{\phi^2-A^2}}\, \sqrt{1-(\del_rt)^2}\ \ .
\ee
For the time variable, we will need to examine the second order equation
of motion: this is difficult in general, but we will discuss this later
in the semiclassical regime.

To gain some intuition for the behaviour of the extremal surface,
let us recall the simple familiar subcase here, of pure $AdS$ with 
no time-dependence: the minimal surface here lies on a constant
time slice. Then comparing with (\ref{phie^fPsi-ansatz}) with all
$t$-exponents vanishing, we have
\be
ds^2 = {R^2\over r^2} (-dt^2+dr^2+dx_i^2)\,,\quad
(\del_rx)^2 = A^2\,{e^f/\phi^{(d_i+1)/d_i} \over \phi^2-A^2}\ ,\quad
S = {V_{d_i-1}\over 4G_{d_i+2}} \int dr\; 
{e^{f/2}\,\phi^{(3-1/d_i)/2}\over \sqrt{\phi^2-A^2}}\,.
\ee
We have\ ${e^f\over \phi^{(d_i-1)/d_i}}=\phi^{2/d_i}={R^2\over r^2}$\ so
this recovers the familiar Ryu-Takayanagi $AdS$ expressions\
$(\del_rx)^2={A^2\over \phi^2-A^2}={A^2r^{2d_i}\over R^{2d_i}-A^2r^{2d_i}}$\
and\ $S \sim {V_{d_i-1} \over G_{d_i+2}} \int
{dr\, \phi^{2}\over\sqrt{\phi^2-A^2}} \sim
{R^{d_i}\,V_{d_i-1} \over G_{d_i+2}} \int
{dr/r^{d_i}\over\sqrt{1-A^2r^{2d_i}/R^{2d_i}}}$\,.
This shows the turning point $r_*$\ (the deepest location till which
the surface dips into the bulk) at
\be
(\del_rx)^2\ra\infty \quad\ra\quad A=\phi_*={R^{d_i}\over r_*^{d_i}}\ ;\qquad
l\sim r_* \sim {R\over A^{1/d_i}}\ .
\ee
The last scaling relation (upto numerical factors) between the width
$l$ and the parameter $A$ arises from using the above expressions in
the width boundary condition in (\ref{RTsurf-bdycndns}). We see that
as the strip width $l$ increases, $r_*$ increases so the surface is
dipping deeper into the bulk interior (and correspondingly $A$ decreases).

In the present cosmological background, for a strip with some fixed
width, the surface begins to dip into the bulk radial direction, which
stops at the turning point $r_*$ where $(\del_rx)^2\ra\infty$.
From (\ref{extSurf-Xf}), we have
\be\label{trngpt-tr}
(\del_rx)^2\ra\infty\ \ \Rightarrow\ \ \left({e^f\over \phi^{(d_i+1)/d_i}}
   \, {  \big(1-(\partial_{r}t)^{2}\big)
  \over \phi^2-A^2}\right)\Big|_{r_*} \ra\infty \ .
\ee
In this case, $\phi, e^f$ also have time-dependence besides
$r$-dependence. However we gain some intuition from looking at the
limit of small strip subsystems in a region far from strong
time-dependence. In this case the extremal surface can be expected
to behave somewhat similar to the $AdS$ case so the turning point
will be at\ $\phi_*=A$. More pertinently $\phi$ is nonvanishing
and $\del_rt\ll 1$, so the only solution to the turning point
equation (\ref{trngpt-tr}) is $\phi_*=A$. As we now increase the
strip subsystem size, the surface becomes ``bigger'' and dips
further into the bulk, but continues to exist since $\phi$ is
continuous. Thus this branch of extremal surfaces that is continuously
connected to the $AdS$-like branch has turning point
\be\label{trngpt-tr-2}
(\del_rx)^2\ra\infty\quad\ra\quad
A = \phi_* = {t_*\,\over r_*^{|m|}}\ ,\qquad\qquad  t_*\equiv t(r_*)\ .
\ee
We have suppressed the $AdS$-like lengthscale and used the scaling
form\ $\phi=t^kr^m$ in (\ref{phie^fPsi-ansatz}) alongwith the
universal relation $k=1$\ (\ref{univSing}), as well as the fact
that $m=-|m|<0$ for the generic cosmological background, as noted
in \cite{Bhattacharya:2020qil}.\ $m<0$ reflects the transverse area
(which is the dilaton in the 2-dim description) growing towards the
boundary $r\ra 0$. For instance the examples (\ref{flat-2d}),
(\ref{AdSDK-2d}), (\ref{hvL-2d}), reviewed earlier exhibit these
features explicitly.
\begin{figure}[h] 
\hspace{2pc}
\includegraphics[width=10pc]{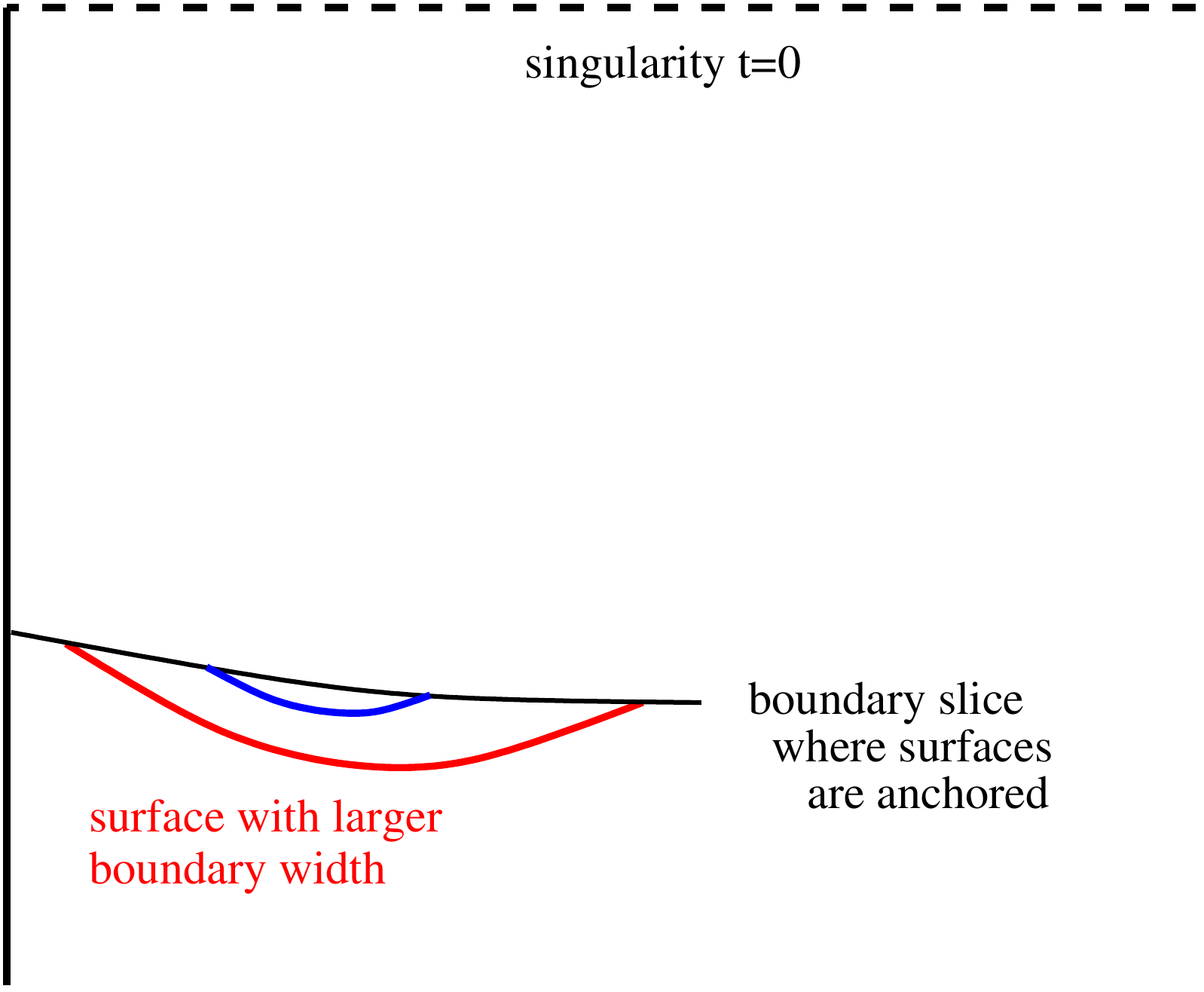}
\hspace{3pc}
\begin{minipage}[b]{24pc}
\caption{{ \label{cosRT1}
    \footnotesize{Cartoon of extremal surfaces in $AdS$ Kasner, \newline
      anchored on a boundary time slice (black curve). For small \newline
      width (blue), the surface stays close to the boundary, while \newline
      for large size (red), the surface dips deeper into the bulk. \newline
      These bend away from the singularity (dotted line). \newline
      }}}
\end{minipage}
\end{figure}

On the face of it, the factor ${e^f\over \phi^{(d_i+1)/d_i}}$ in
(\ref{trngpt-tr}) contains a factor of $t^{-\#}$ and so appears to
lead to a distinct turning point localized at the singularity $t_*=0$
as well. However this branch of extremal surfaces appears disconnected
from the branch that is continuously connected to the $AdS$-like
branch in the region far from the singularity. If such a branch with
$t_*=0$ exists it can only exist in the limit of infinite strip width,
\ie\ the IR limit, with no way to move away from $t_*=0$ locus (since
there is no parameter like $A$ that allows deforming): this implies
it is inaccessible from the classical region far from the
singularity. However the region near the singularity at $t=0$ is a
region where quantum gravity effects must be strong: a classical
RT/HRT extremal surface localized there with no way to deform away to
a well-defined classical region is unreliable. For this
reason, we discard this branch of possible extremal surfaces.  Finally
for $(\del_rx)^2>0$ to be well-defined, we must have\ $(\del_rt)^2<1$
so $|\del_rt|$ is bounded. For small strip width, the extremal surface
lies on an almost-constant time slice \ie\ $(\del_rt)^2\ll 1$.
These suggest that there cannot arise any divergence in
(\ref{trngpt-tr}) from the $(1-(\del_rt)^2)$ term. The condition
$(\del_rt)^2<1$ is consistent with the surface being spacelike
everywhere for our boundary conditions (\ref{RTsurf-bdycndns})\
(unlike \eg\ \cite{Narayan:2012ks} where the anisotropy induced by the
energy flux implied that for the strip orthogonal to the flux there is
a phase transition in the surfaces).
Overall these arguments pin down the condition (\ref{trngpt-tr-2})
as the relevant one for the turning point of the extremal surface
(\ref{extSurf-Xf}) of interest.

Going with the reasonable assumption that the time direction is not
doing anything singular, as described above, we will now
examine the scaling of the width with $t_*, r_*, A$. It is instructive
to focus on the $AdS$ Kasner spacetime (\ref{AdSDK-2d}) for this purpose:
(\ref{extSurf-Xf}) then gives
\be\label{extSurf-AdSKas}
(\del_rx)^2 = A^2 \left({1\over t^{2/d_i}}\right)
   {1-(\partial_{r}t)^{2} \over {t^2\over r^{2d_i}}-A^2}\ ,\qquad
S = {V_{d_i-1}\over 4G_{d_i+2}} \int dr\;
\Big({t^{2-1/d_i}\over r^{2d_i}}\Big)
{\sqrt{1-(\del_rt)^2}\over \sqrt{{t^2\over r^{2d_i}}-A^2}}\ .
\ee
These expressions can be used to explicitly see our general statements
earlier. The spatial width condition (\ref{RTsurf-bdycndns}) in this case
gives
\be\label{lr*t*A}
   {l\over 2} = \int_0^{r_*} dr\, (\del_rx)
   = A \int_0^{r_*} dr\ {e^{f/2}\over \phi^{(d_i+1)/2d_i}}\,
   \sqrt{{1-(\partial_{r}t)^{2}\over \phi^2-A^2}}\,
   =\, r_*\int_0^1 {du\over t^{1/d_i}}\,
   {\sqrt{1-(\del_rt)^2}\over \sqrt{(\phi/\phi_*)^2-1}}\ ,
\ee
where $\phi=tr^{-d_i}$ and using (\ref{trngpt-tr-2}). Now the integral
has no nontrivial scale dependence: it has been absorbed into the
$r_*$ factor outside. This gives the scaling, using (\ref{trngpt-tr-2}),
\be\label{lr*t*A-2}
l\ \sim\ r_*\ ; \qquad 
A = {t_*\over r_*^{d_i}} \sim {t_*\over l^{d_i}} \ .
\ee
This fits the expectation that as the width $l$ increases, the surface
dips deeper into the bulk so the radial turning point $r_*$
increases. The dip in the time direction is mild at least when the
surface is anchored on a time slice far from the singularity at $t=0$:
in this case the surface almost lies entirely on a constant time slice
$t\sim t_0\gg 0$ so $t_*\sim t_0$ as well (as we discuss later). In
this regime, $A\sim {1\over l^{d_i}}$\,.  The $u$-integral in
(\ref{lr*t*A-2}), with each term positive, gives a positive numerical
factor.  The scaling $t_*\sim Ar_*^{d_i}$ suggests that increasing
$r_*$ implies increasing $t_*$. Since in this entire discussion, we
restrict to one side of the singularity (the past), the range of the
time variable $t$ is $t\equiv |t|\geq 0$, so increasing $t_*$ means
bending away from the singularity at $t=0$.  Similar observations were
noted in \cite{Engelhardt:2013jda} in a different context.

All the above arguments are reasonable as long as we are in the
semiclassical regime far from the singularity: but they do not
pin down $t_*$.
More information on the time behaviour, \ie\ the $t(r)$ function,
is obtained by analysing the equation obtained from extremizing
(\ref{EEtxr-areaFnal}) with respect to the $t$-variable: this gives
the $t$-equation of motion\ 
\begin{multline}\label{teqm}
\frac{d}{dr}\ \bigg\{ \frac{\phi\ e^{f} \ (\partial_{r}t)}{\phi^{(d_{i}+1)/d_{i}}\ \sqrt{{e^f\over \phi^{(d_i+1)/d_i}}\big(1-(\partial_{r}t)^{2}\big)+(\partial_{r}x)^{2}}}\bigg\}\ 
+ \ {\dot\phi}\ \sqrt{{e^f\over \phi^{(d_i+1)/d_i}}\big(1-(\partial_{r}t)^{2}\big)+(\partial_{r}x)^{2}} \\
+ \frac{\phi\ (1-(\partial_{r}t)^{2})\ e^{f}\ \big\{\ \dot{f} - \frac{(d_{i}+1)}{d_{i}}\ \frac{{\dot\phi}}{\phi}\ \big\}}{2\phi^{(d_{i}+1)/d_{i}}\ \sqrt{{e^f\over \phi^{(d_i+1)/d_i}}\big(1-(\partial_{r}t)^{2}\big)+(\partial_{r}x)^{2}}} = 0\ , 
\end{multline}
with ${\dot\phi} = \frac{\partial \phi}{\partial t}$ etc.
Using the conserved quantity (\ref{extSurf-Xf}) for the $x$-variable, this
simplifies. Then specialising to the $AdS$ Kasner case, we obtain
\begin{multline}\label{t-EOM}
\frac{d}{dr}\bigg\{ \frac{\partial_{r}t}{\sqrt{1-(\partial_{r}t)^{2}}}\ \frac{t^{-1/d_{i}}}{r^{d_{i}}}\ \sqrt{t^{2}-r^{2d_{i}}A^{2}}\bigg\}\ +\ \frac{t^{(d_{i}-1)/d_{i}}}{r^{d_{i}}}\ \sqrt{\frac{1-(\partial_{r}t)^{2}}{t^{2}-r^{2d_{i}}A^{2}}} \\ -\ \frac{1}{d_{i}}\ \frac{1}{t^{(d_{i}+1)/d_{i}}\,r^{d_i}}\ \sqrt{(1-(\partial_{r}t)^{2})(t^{2}-r^{2d_{i}}A^{2})} = 0\ .
\end{multline}
This gives, with $t'\equiv \del_rt$,
\be\label{del^2rt}
(1-t'^2)\left(d_i^2t'+{r(t^2-A^2r^{2d_i}) \over t^3} - {d_ir\over t}\right)
-{(t^2-A^2r^{2d_i}) d_ir t'' \over t^3} = 0\ .
\ee
We have suppressed the Kasner scale $t_K$ mentioned after
(\ref{AdSDK-2d}): reinstating this shows that $A$ appearing above is
really $At_K$, so that each term above is dimensionless (we have
suppressed the $AdS$ scale $R$: reinstating that we can rescale $t, r$
by $R$, and then the above statement on each term being dimensionless
continues to hold).  Now note that we have written (\ref{del^2rt}) to
emphasise that in the region far from the singularity at $t=0$, we
have large $t$ so we can analyse this equation in detail. Here, as
stated before, we expect that the surface will have only mild time
dependence, lying almost on a constant time slice, so $\del_rt\ll
1$. At the turning point, from (\ref{trngpt-tr}), (\ref{trngpt-tr-2}),
we have\ $\phi_*={t_*\over r_*^{d_i}}=A$ and $(\del_rx)^2\ra\infty$.
Then the nonzero terms in (\ref{del^2rt}) give
\be\label{t'*}
t'_* = {r_*\over d_i\,t_*} > 0\qquad\ra\qquad  {dr\over dx}\Big|_* = 0\ ,
\qquad {dt\over dx}\Big|_* = {t'_*\over (\del_rx)_*} = 0\ .
\ee
Thus the extremal surface is at a $t$-maximum at the turning point
$(r_*,t_*)$. Further in its neighbourhood, we have
\be\label{t'*-2}
r<r_*\,,\ \ t'_*>0 \quad\ra\quad t(r) \sim t_* + t'_*\,(r-r_*) < t_*\ ,
\ee
verifying that $t_*$ is a local maximum (we recall that the range of
$t\equiv |t|>0$ so $t$ increasing is going away from $t=0$).
\begin{figure}[h] 
\hspace{2pc}
\includegraphics[width=6.5pc]{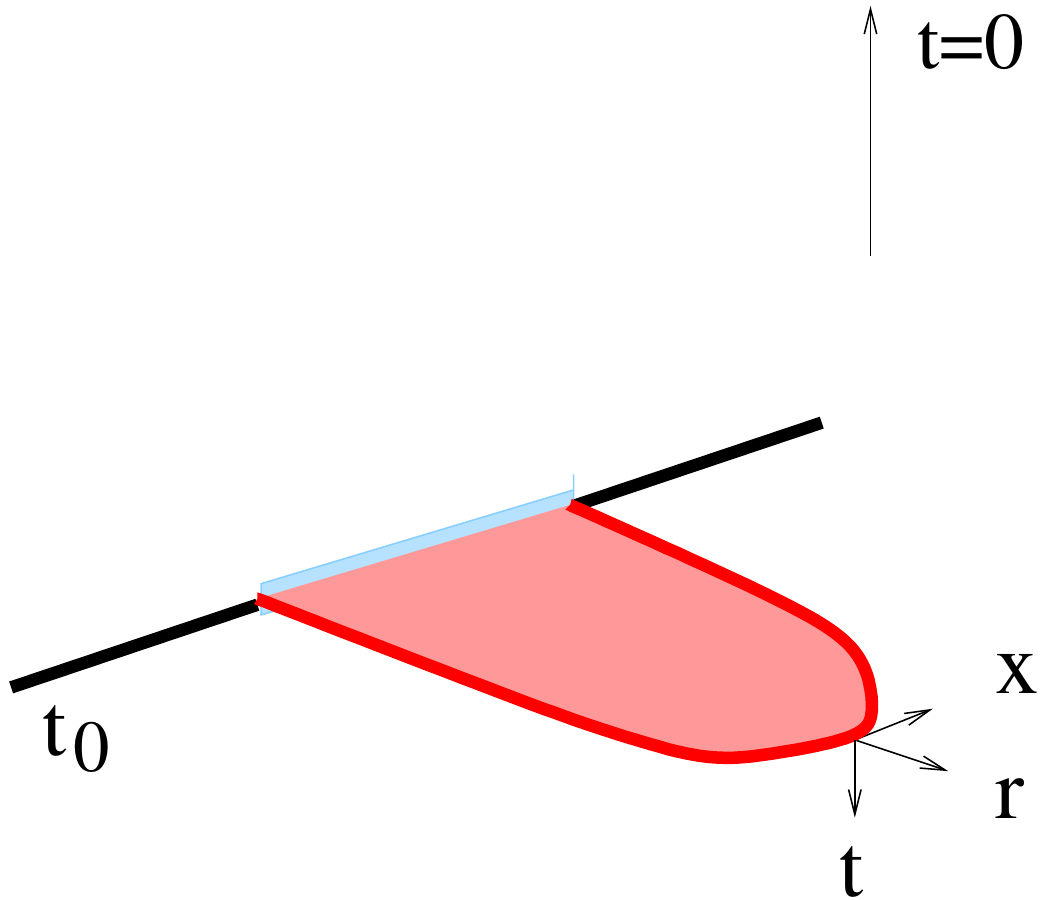}
\hspace{3pc}
\begin{minipage}[b]{24pc}
\caption{{ \label{cosRT3}
    \footnotesize{Cartoon of the local geometry of the extremal surface 
      \newline near the turning point $(r_*, t_*)$. The surface is anchored
      on a \newline time  slice $t_0$ far from the singularity at $t=0$
      \ (not shown). \newline It dips in the direction away from the
      singularity. 
      }}}
\end{minipage}
\end{figure}
This suggests the global condition $t_*>t_0$ at least for sufficiently
small size (sufficiently small $r$), although strictly speaking that
is not implied. To gain insight into this, we can employ perturbation
theory in (\ref{del^2rt}) to study the $t(r)$ function in the region
far from the singularity. Since the time dependence there is expected
to be small, we can take $t'\ll 1$ and therefore approximate the
equation (\ref{del^2rt}) by
\be\label{del^2rt-2}
d_i^2t^3t' + r(t^2-A^2r^{2d_i}) - d_ir t^2 - (t^2-A^2r^{2d_i}) d_ir t'' = 0\ ,
\ee
dropping $t'^2$ in the first $(1-t'^2)$ factor. Now a power series
ansatz for $t(r)$ gives
\be
t(r) = t_0 + \sum_{n\in \BZ^+} c_n r^n \quad\ra\quad c_n\sim {1\over t_0^\#}\ .
\ee
We have $t(0)=t_0$ as a boundary condition for anchoring the surface at
the boundary. 
Sticking this series in (\ref{del^2rt-2}) can be done numerically in
Mathematica, and the $c_n$ can be solved for iteratively: they scale
inversely with $t_0$.
This shows for $r\leq r_*\lesssim t_0$ that $t(r)$ is indeed almost
constant, with only mild variation. For example in $AdS_5$ Kasner
($d_i=3$), we find (see Appendix A for more details, specifically
(\ref{AdS5-t(r)-A}))
\be\label{AdS5-t(r)}
t(r) = t_0 + {1\over 12 t_0}\,r^2 - {1\over 432 t_0^3}\,r^4
+ {1\over 7776 t_0^5}\,r^6
+ \Big({A^2\over 160 t_0^3} - {17\over 7776\cdot 240 t_0^7}\Big)\,r^8 + \ldots
\ee
with higher order terms further suppressed.
The first few terms scale as ${1\over t_0^{n-1}}$ so they are manifestly
subleading in the regime $r_*\lesssim t_0$. The higher terms beginning
with $r^8$ contain $A$ in their coefficients. Recall now that $A$ here
is really $At_K$, reinstating the Kasner scale $t_K$ as mentioned after
(\ref{del^2rt}). Thus $At_K=t_k{t_*/t_K\over r_*^3}\sim {t_0\over r_*^3}$
since $t_*\sim t_0$. This means the term containing $A^2$ in
the $r^8$-term is of the form\ ${t_*^2\, r^8\over 160 r_*^6\,t_0^3} \sim
{r_*^2\over 160 t_0}\,{r^8\over r_*^8}\ll t_0$ since $t_*\sim t_0$
and $r\leq r_*\lesssim t_0$\,: so this term is also suppressed compared
with the leading $t_0$ term. Now between the two terms in the $r^8$
coefficient, we see that\
${A^2\over t_0^3}\sim {t_*^2\over r_*^6\,t_0^3}\sim {1\over r_*^6\,t_0}$
dominates over ${1\over t_0^7}$\,. This is also true at higher orders
where there are further terms containing ${A^{2k}\over t_0^\#}$\,.
Analysing this further and retaining only the dominant terms in
each $r^k$-coefficient leads to a new series (see Appendix A).
Evaluating this at the turning point $r_*$ gives (\ref{t(r)-sub}):
thus we manifestly see that
\be\label{t*>t0}
t_* = t(r_*) > t_0\ ,\qquad\qquad r_*\lesssim t_0\ ,\qquad
\ee
In other words, the surface bends away from the singularity at $t=0$,
at least if anchored in the reliable far-region with $t_0\gg 0$. Using
this it can be seen that $t'^2\ll 1$ indeed so our approximation of using
(\ref{del^2rt-2}) instead of (\ref{del^2rt}) is justified in this regime.
Similar power series and results arise in $AdS_4$- and $AdS_7$-Kasner.
Thus overall, the surface function $t(r)$ starts at $t=t_0$ at the
boundary $r=0$ and then grows, reaching a maximum value at the turning
point $t=t_*$ bending away from the singularity. Then the surface
turns around and returns to the boundary (joining the other end of the
strip subsystem). This is depicted in Fig.~\ref{cosRT3}.

The IR limit where the strip width is large (using (\ref{lr*t*A-2}))
is ${t_*\over r_*^{d_i}}=A\sim {1\over l^{d_i}}\ra 0$. In this limit, 
(\ref{del^2rt-2}) becomes\
$d_i^2t^3t' + rt^2 - d_ir t^2 - t^2 d_ir t'' = 0$. Again analysing via
a power series in Mathematica gives in $AdS_5$ Kasner ($d_i=3$), we find
\be
t(r) = t_0 + {1\over 12\,t_0}\,r^2 - {1\over 432\,t_0^3}\,r^4
+ {1\over 7776\,t_0^5}\,r^6
- {17\over 1866240\,t_0^7}\,r^8 +
 {247\over 335923200\,t_0^9}\,r^{10} + \ldots
\ee
The series here is more delicate since the surface really has
$r_*\ra\infty$ (dipping into the bulk fully) so the entire $r$-series
is important. The limit $A\ra 0$ requires $A\lesssim {1\over t_0^2}$
comparing with the scale $t_0$: this requires
${t_0\over r_*}\lesssim 1$\,. Thus the IR limit here is 
\be\label{t*r*-A=0}
r_*\ra\infty\ ,\quad t_0\ra\infty\ , \qquad\quad {t_0\over r_*}\lesssim 1\ ,
\ee
which is the reliable semiclassical regime far from the singularity.
In this regime the series defining the time behaviour of the surface
continues to be well-defined, albeit delicate: the surface is anchored
on a slice far from $t=0$ so although it dips deep into the bulk, its
time dependence is mild with $t'^2\ll 1$ everywhere. This then shows
that\ $t_*>t_0$\ for ${t_0\over r_*}\sim 1$\,:\ numerically it can be
checked that for ${r_*\over t_0}\lesssim 3$ the $t(r)$ series above
continues to satisfy $t_*>t_0$, \ie\ the extremal surface bends in
the direction away from the singularity. As ${r_*\over t_0}$ increases
further (keeping the limit (\ref{t*r*-A=0}), $t_0$ becomes smaller
and it can be seen numerically that the series above violates $t_*>t_0$:
however in this regime, it can also be seen that $t'$ is increasing
so the analysis is breaking down: this occurs as we move the anchoring
surface in the direction of the singularity, which becomes unreliable
(not surprisingly).

The conditions (\ref{t'*}), (\ref{t'*-2}), on the local geometry in
the neighbourhood of the turning point do not depend on $A$, so in
particular they also apply in this IR limit.  This is consistent with
the spacelike condition being preserved here for generic strip
size. Overall we see that such classical RT/HRT extremal surfaces
exist for generic strip size.  It is interesting that such a power
series analysis works, since (\ref{del^2rt-2}) and its $A=0$ limit are
still complicated nonlinear equations: as we depart from the large
$t_0$ semiclassical regime and move the anchoring surface toward the
singularity, it is unclear if this can be analysed meaningfully. We
have mainly analysed the $AdS$ Kasner spacetime here for the time
behaviour: however the techniques should be applicable to more
general cosmologies of the kind described earlier.

The IR limit of large size $l$ can be probed in greater detail by
quantum extremal surfaces as we will see in the next section. The
findings there are consistent with the classical RT/HRT analysis here,
but constrain $t_*, r_*$ further, owing to the bulk matter entropy
contribution.

Finally, we now make some general statements based on energy conditions.
Firstly for (\ref{del^2rt}) as well as (\ref{extSurf-AdSKas}) and
(\ref{lr*t*A}) to be well-behaved, we have seen that\ $(\del_rt)^2<1$\
which follows from requiring reality, and also $\phi\geq\phi_*=A$. The
first condition is expected intuitively as we have seen.  The second
condition implies $\phi={t\over r^{d_i}}$ decreases till it becomes
$\phi_*$, \ie\
\be
\phi\geq\phi_*\ \ \ra\ \
   {t(r)\over r^{d_i}} \geq {t_*\over r_*^{2d_i}}\qquad\ra\qquad
   \del_r \Big({t(r)\over r^{d_i}}\Big) \leq 0\ \ \Rightarrow\ \
   \del_rt\leq {d_it\over r}\ .
\ee
The derivative condition follows from assuming monotonicity (with $r=0$
the boundary). This constrains the behaviour of the $t(r)$ function.
More generally, this is reminiscent of null energy conditions, and the
dilatonic c-function in \cite{Kolekar:2018chf}. Using the equations
(\ref{2dimseom-EMD0-Psi}) in the 2-dim background, the NEC gives\
$-n^\mu n^\nu\nabla_\mu\nabla_\nu\phi
= g^{tt}\nabla_t\nabla_t\phi-g^{rr}\nabla_r\nabla_r\phi\geq 0$,
\ie\ \
$-e^f(\del_t^2\phi-\del_tf\del_t\phi + \del_r^2\phi-\del_rf\del_r\phi)\geq 0$\
which simplifies to\ $-(\phi''-f'\phi'-{(d_i-1)/d_i\over t\,r^{d_i}})\geq 0$\
using (\ref{AdSDK-2d}).
Restricting to the extremal surface we have\ $\phi={t(r)\over r^{d_i}}$
and $f={d_i-1\over d_i}\log t(r) - (d_i+1)\log r$ which gives\
$t'' \leq {d_i-1\over d_i}\, {1+t'^2\over t}$\,. These general
conditions appear consistent with the earlier discussions: although
we have not used these much these considerations may provide
interesting information in general cosmologies.

\section{Quantum extremal surfaces}

We will now study quantum extremal surfaces (QES) in the 2-dim
cosmologies in \cite{Bhattacharya:2020qil} obtained by dimensional
reduction from various higher dimensional theories, towards
understanding the cosmological, Big-Crunch, singularities present
here, in part inspired by the exciting findings in
\cite{Penington:2019npb,Almheiri:2019psf}.  The relevant 2-dim fields
here are the dilaton $\phi$, the 2-dim metric $e^f$ and the extra
scalar $\Psi$.  The scalar $\Psi$ is essential for nontrivial dynamics
and essentially drives the singularity: however since the spacetime
already contains the effects of the scalar we will assume that the
scalar excitations are on the same footing as other bulk matter.  This
is equivalent to assuming that the effects of $\Psi$ have been
subsumed into their backreaction on the geometry so using the
background spacetime is adequate. Thus the bulk matter entropy
$S_{bulk}$ will be assumed to contain contributions from the scalar
$\Psi$ as well, which will not be treated separately. Towards putting
this on firmer footing, it is important to understand the bulk
entanglement entropy for scalars such as $\Psi$ with a dilaton
coupling in the action (\ref{actionXPsiU}): we will leave this for the
future.

Consider an observer $O$ at some location $(t_0,r_0)$ moving in time
in the spacetime background. In the time-dependent case, if he/she is
far away from regions such as the Big-Crunch singularity, the time
dependence is slow and it is reasonable to imagine that the ambient
matter in the observer's neighbouring patch is in its ground
state. Now we ask what entanglement he/she sees: say the QES is at some
location $(t,r)$. We will use $(t_*,r_*)$ to refer to the QES solution
to extremization of the generalized entropy,
\bea\label{Sgen0}
S_{gen} &=& {\phi\over 4G_2} + S_{bulk} \nn\\
&=& {\phi\over 4G_2} +
{c\over 12} \log \left( \Delta^2\ e^{f}\big|_{(t,r)} \right) +\ \ldots 
\eea
This is the classical area (dilaton) piece along with the subleading
entropy of bulk matter. $\Delta^2$ is the flat spacetime interval
between the observer $O$ and the QES,
\be\label{Delta2}
\Delta^2 = r^2 - (t-t_0)^2\ .
\ee
We focus on the observer $O$ located at the boundary $r=0$ since all
the backgrounds we discuss have a holographic dual interpretation.
We have written the expression (\ref{Sgen0}) along the lines of the
discussions in \cite{Almheiri:2019psf}: in particular the effects of
the curved spacetime appear entirely through the conformal factor
$e^f$ at the QES endpoint of the interval (see Appendix B for a very
brief recap).  Above, we have only explicitly retained terms that are
relevant for the QES extremization. So we have omitted terms
containing the ultraviolet cutoffs and the warp factor
$e^f|_{t_c,r_c}$ at the boundary: the latter can be partially absorbed
into the UV cutoffs.  A useful resource for these QES calculations in
time-independent cases is \cite{MahajanTalk}.

\medskip

\noindent Several comments are in order on the generalized entropy
(\ref{Sgen0}) in our description and use: while some of these are
also features of previous applications of the generalized entropy,
some are specific to our context as will be clear in what follows.
\begin{enumerate}
  \item{The only boundary subregions that make sense in these
reduced 2-dim bulk theories are the whole space from the higher
dimensional point of view. So the leading term is the transverse area
of the full space in the higher dimensional theory, which is the 2-dim
dilaton (in Planck units).  From the expression (\ref{EEtxr-areaFnal})
for the RT/HRT surface in the higher dimensional theory, this leading
classical term can be seen to arise as
\be
S_{gen}^{cl}\ \sim\ {V_{d_i-1}\over 4G_{d_i+2}} \int dr\, \phi\,
\sqrt{(\partial_{r}x)^{2}}\ \sim\ {V_{d_i-1}\over 4G_{d_i+2}} \int dx\, \phi
\ \sim\ {\phi\over 4G_2}\ .
\ee
using ${1\over G_2}\sim {V_{d_i}\over G_{d_i+2}}$\,.
In the 2-dim theory, the extremal surface is just a point in the 2-dim
spacetime, the entire transverse part of the higher dim extremal surface
wrapped. The QES location $(t_*,r_*)$ after extremization of
the generalized entropy (\ref{Sgen0}) is roughly speaking analogous to
the turning point in the classical RT/HRT analysis, and the 2-dim
discussion pertains to the IR limit there.\\
In time-independent situations, the boundary subregion can be taken to
be a point on the boundary $r=r_c$ on some constant time slice, and
then the extremal surface is a point lying at some spatial location
$r=r_*$ on that slice. We will describe some examples of this below. }

\item{In general we will assume that the bulk matter is a 2-dim
  conformal field theory. Then the bulk entanglement entropy
  \cite{Faulkner:2013ana,Engelhardt:2014gca} of quantum matter fields
  in the bulk subregion enclosed by the extremal surface and the
  boundary subregion can be described by the Calabrese-Cardy
  expression \cite{Calabrese:2004eu,Calabrese:2009qy} for a 2-dim
  $CFT$ in flat space, along with the modifications from the conformal
  transformation to the curved 2-dim space \cite{Almheiri:2019psf}.
  The interval has endpoints defined by the extremal surface at
  $(t,r)$ and the boundary $(t_0,r_c)\sim (t_0,0)$: thus $S_{bulk}$
  has been obtained using the rules of boundary CFT, with a single
  twist operator at the QES endpoint (some details appear in Appendix
  B). For the cosmological spacetimes, nontrivial time dependence
  arises from the interval but mainly from the conformal
  transformation, which as we will see is nontrivial.\\
  We have also assumed in writing this expression that
\be
1\ll c \ll {1\over G}\ ,
\ee
\ie\ the matter CFT central charge is sufficiently large to give
nontrivial subleading contributions to the generalized entropy, but
not too large that it backreacts and wrecks the classical geometry
(and thereby the classical entanglement term).\\
In the higher dimensional theory, the bulk entropy contribution
\cite{Faulkner:2013ana,Engelhardt:2014gca} is
in general difficult to calculate: we resort to the effective 2-dim
theories where $S_{bulk}$ can be approximated by 2-dim CFT entanglement
entropy.}

\item{In writing $S_{bulk}$ we have assumed that the quantum matter
  fields are in a pure state, and for simplicity we have assumed the
  ground state of the CFT for most of our analysis. This is reasonable
  if the background is time-independent, or has slow time variation.
  In the cosmological cases we discuss, this form of $S_{bulk}$ is
  reasonable in the semiclassical spacetime region far from the
  singularity where the time variations are not significant. Since the
  warp factor $e^f$ contains time-dependence, $S_{bulk}$ contains
  effects of time evolution: we expect this to be reasonable for mild
  time dependence.

  However in the cosmological spacetimes we discuss, there are global
  Big-Crunch singularities where the entire spacetime becomes
  vanishingly small with the conformal factor $e^f$ going to zero:
  this suggests a singularity arising from $\log e^f$.  Physically one
  might imagine that the severe time-dependence would lead to the bulk
  matter going to some excited state, perhaps severely excited near
  the singularity. This suggests a breakdown of $S_{bulk}$, and in
  fact the entire semiclassical approach in these techniques which are
  unceremoniously being extrapolated to a region with large quantum
  gravity effects. As it turns out, our analysis appears
  self-consistent in the sense that the quantum extremal surfaces end
  up being localized in the semiclassical spacetime region far from
  the singularity.  We will comment on these further after we discuss
  the analysis.}

\end{enumerate}

\subsection{Some time-independent backgrounds}

Before studying the cosmological backgrounds, we will first study some
time-independent backgrounds to gain some intuition and experience for
the above generalized entropy (\ref{Sgen0}) and the resulting quantum
extremal surfaces.  With no time-dependence, all time slices are
equivalent so we can set $t=t_0$, \ie\ the QES lies on the same time
slice as the observer.  This is of course borne out in our experience
with entangling surfaces in time-independent backgrounds in higher
dimensional holography.

\bigskip

\noindent \underline{\bf $AdS_2$}

Here we have $ds^2={1\over r^2}(-dt^2+dr^2)$ with $\phi={\phi_r\over r}$
and the generalized entropy (\ref{Sgen0}) setting $t=t_0$ becomes
\be
S_{gen} = {\phi_r\over 4G}{1\over r} + {c\over 6}\log \Big(r\,{1\over r}\Big)\ ;
\qquad \del_rS_{gen} \sim -{\phi_r\over r^2}\ra 0\ .
\ee
We have retained only terms relevant for the extremization. We see
that the warp factor at the $r$-endpoint cancels the $r$-dependence of
the interval entanglement, giving just the classical piece. Thus the
extremization gives the second expression above so\ the solution to
extremization is\ $r_*\ra\infty$.
Thus the entanglement wedge \cite{Czech:2012bh,Wall:2012uf,Headrick:2014cta}
defined as the bulk domain of dependence
of the QES (the part of the spacetime causally connected to the QES
at $r_*\ra\infty$) is the entire Poincare wedge as expected.\ \
See also \cite{MahajanTalk} for discussions on this.

\bigskip

\noindent \underline{\bf $AdS_D$ reduction}

The higher dim $AdS_D$ space with $D=d_i+2$ is\
$ds^2_{AdS_{d_i+2}}={1\over r^2} (-dt^2+dr^2)+{1\over r^2} dx_i^2$ and under
reduction (\ref{redux+Weyl}) we obtain the 2-dim background (suppressing
the $AdS$ scale)
\be\label{AdSDred}
\phi = {1\over r^{d_i}}\ ,\qquad 
ds^2 = {1\over r^{d_i+1}} (-dt^2+dr^2)\ .
\ee
Some aspects of such generic 2-dim dilaton gravity theories have been
discussed in \cite{Narayan:2020pyj}. Now (\ref{Sgen0}) gives
\be\label{SgenAdSDred}
S_{gen} = {\phi_r\over 4G}\,{1\over r^{d_i}}
+ {c\over 6} \log \left( {r\over r^{(d_i+1)/2}} \right)\quad\ 
\Rightarrow\quad\ \del_rS_{gen} = -{d_i\phi_r\over 4G\, r^{d_i+1}}
- {c\over 6}\,\Big({d_i-1\over 2}\Big)\,{1\over r} = 0\ .
\ee
We see that both terms are are of the same sign since $c>0$ and $d_i>1$.
Thus the solution is again\ $r_*\ra\infty$ for the location of the QES:
this again leads to the entire Poincare wedge which is the expected
answer (also in the higher dimensional point $AdS_D$ when the subsystem
becomes the whole space).
Note that we are using the 2-dim metric as the Weyl transformed one
(\ref{redux+Weyl}) in (\ref{AdSDred}) above: this was found to be
consistent in \cite{Narayan:2020pyj} in holographic discussions (\eg\
the stress tensor).

It is to be noted that we have written $S_{bulk}$ using the rules of
boundary CFT since the effective space is the half-line with one end
of the interval at the boundary $r=0$. It is instructive to compare
this with the discussion of islands in \eg\ \cite{Almheiri:2019yqk},
where a flat region was appended beyond the boundary $r=0$ of an
$AdS_2$ region: in this case, the generalized entropy takes the
form\
$S_{gen}\sim {\phi_r\over 4G}{1\over r} + {c\over 6}\log ((r+r')^2\,{1\over r})$.
The interval in question has endpoints $r\in AdS_2$ and $r'$ in the
flat space region beyond the boundary: the warp factor at the $r'$ end
does not contribute since it is trivial in that flat region.
Both $r, r'>0$ in this parametrization: the space is not a half-line
now.  Let us set $r'\sim 0$ for simplicity. Then extremizing gives\
$-{\phi_r\over 4G}{1\over r^2} + {c\over 6}{1\over r} = 0$\,: the 
competition between the two terms leads to a finite value
$r_*\sim {\phi_r\over Gc}$ for the QES location. In the case
(\ref{SgenAdSDred}) above we see that the argument of the logarithm
in the bulk entropy arises differently and $r_*\ra\infty$ with no island.

One way to understand this is in terms of the violation of the
Bekenstein bound, as discussed in \cite{Hartman:2020khs}: if the
classical dilatonic term is overpowered by the subleading bulk
entropy contribution, we may expect islands. To see this, note that
(\ref{SgenAdSDred}) can be recast as
\be
S_{gen} = {\phi\over 4G} + {c\over 6}\,{d_i-1\over d_i}\,\log\phi\ ,
\ee
with a relative plus sign in the two contributions, again retaining
only terms relevant for extremization\
(in greater detail, putting the $AdS$ scale and the UV cutoff scales
back, the bulk entropy term is\
$\log (\phi^{(d_i-1)/d_i} {R^2\over\epsilon_1\epsilon_2})$\,: the
argument becomes $O(1)$ when $\phi$ is sufficiently small, at large $r$).
As long as $\phi$ is not too small, $\log\phi$ will always be
subdominant to the classical area term $\phi$ and the Bekenstein
bound will not be violated: thus the extremization leads to
$\del_r\phi=0$ giving $r_*\ra\infty$ which is the entire Poincare
wedge, with no islands.
If one could somehow entangle the bulk matter on the interval with
some other region, this may lead to $S_{bulk}$ increasing and
overpowering the classical area contribution: this is what appears to
be happening in the example above from \cite{Almheiri:2019yqk}, as
well as various cases discussed in \cite{Hartman:2020khs}.


\subsection{2-dim cosmologies and quantum extremal surfaces}

Now we will study quantum extremal surfaces in the 2-dim cosmological
backgrounds reviewed earlier. We focus first on the 2-dim cosmology
obtained by reduction of the $AdS_D$ Kasner spacetime (\ref{AdSDK-2d}).
With the observer at the boundary and the interval $\Delta^2$ between
the observer and the QES location as in (\ref{Delta2}), we obtain
\bea\label{SgenAdSKas}
&& S_{gen} = {\phi_r\over 4G} {t\over r^{d_i}} + {c\over 12} \log
\left[ \big(r^2-(t-t_0)^2\big)\, e^f\big|_{(t,r)} \right] + \ldots \\ [2mm]
&& \del_rS_{gen} = -{\phi_r\over 4G}\,{d_i\,t\over r^{d_i+1}} +
{c\over 6}\,{r\over r^2-(t-t_0)^2} - {c\over 12}\,{d_i+1\over r} = 0\ ,  \nn\\
    [2mm]
&& \del_tS_{gen} = {\phi_r\over 4G}\,{1\over r^{d_i}} - {c\over 6}\,
    {t-t_0\over r^2-(t-t_0)^2} + {c\over 12}\,{d_i-1\over d_i\,t} = 0\ .
    \label{SgenAdSKas2}
\eea
\begin{figure}[h] 
\hspace{2pc}
\includegraphics[width=6.5pc]{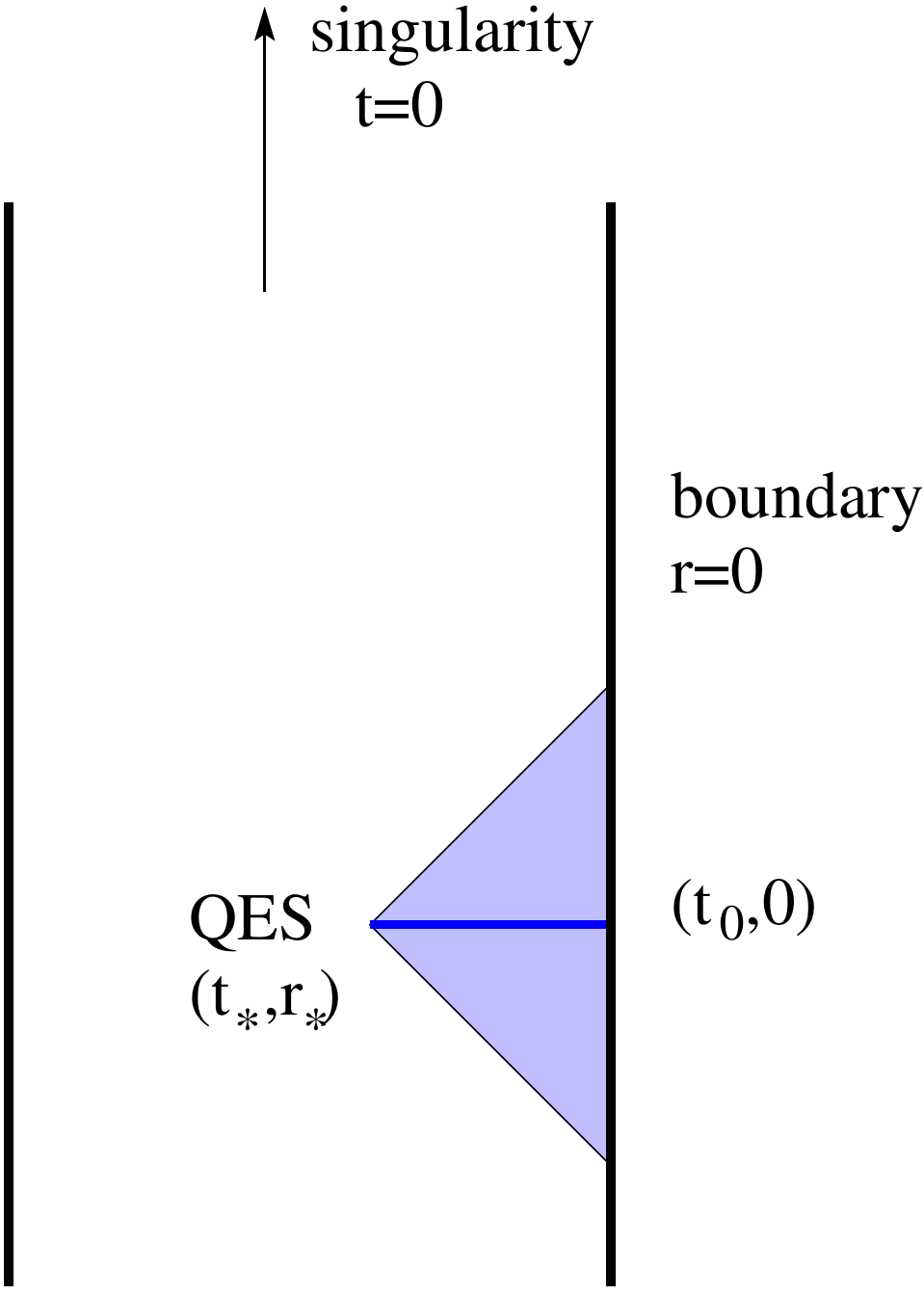}
\hspace{3pc}
\begin{minipage}[b]{26pc}
\caption{{ \label{cosQES}
    \footnotesize{Cartoon of the 2-dim geometry, the holographic
      boundary \newline at $r=0$ and the QES at $(t_*,r_*)$. The solid
      blue line is the spatial interval between the boundary observer
      and the QES. The singularity is \newline at $t=0$ far up (not
      shown). The extremization drives the QES location \newline to
      $t_*\ra\infty$ (far from the singularity) and $r_*\ra\infty$
      (which is the left \newline vertical line).
      }}}
\end{minipage}
\end{figure}    
We are again only retaining terms relevant for extremization.
Some general comments can be made here: assuming as is usually the case
in $AdS/CFT$,  the entanglement wedge lies outside the causal wedge
(deeper in the interior)
\cite{Czech:2012bh,Wall:2012uf,Headrick:2014cta}, we have $\Delta^2>0$.
For the usual parametrization of the bulk, we have $r>0$: this also
implies $\Delta^2>0$ for a real solution (independent of the argument
above).\ Rearranging (\ref{SgenAdSKas2}) gives
\be\label{ExtSgen-r0t0}
{c\over 6}\,{r\over r^2-(t-t_0)^2} =
{\phi_r\over 4G}\,{d_i\,t\over r^{d_i+1}} + {c\over 12}\,{d_i+1\over r}\,,\qquad
{c\over 6}\,{t-t_0\over r^2-(t-t_0)^2} =
{\phi_r\over 4G}\,{1\over r^{d_i}} + {c\over 12}\,{d_i-1\over d_i\,t}\ .
\ee
Since $d_i>1$, the right side is positive always, so we have
\be\label{Delta^2>0}
\Delta^2>0\ \ \ie\ \ (t-t_0)^2<r^2 ,\qquad r>0 ,\qquad t\geq t_0\ .
\ee
$t\geq t_0>0$ means the QES lies on the same time slice as or beyond
the observer, further away from the singularity.\\
If we assume that the QES lies on the same time slice as the observer
\ie\ $t=t_0$, equivalently that the QES is maximally spacelike separated
from the observer, we obtain from (\ref{SgenAdSKas})
\bea\label{SgenAdSKas-t=t0}
t=t_0:\qquad\quad
S_{gen} = {\phi\over 4G} + {c\over 6}\,{d_i-1\over d_i} \log\phi\ ,&& \quad
\phi={\phi_r\over 4G}\,{t\over r^{d_i}}\ ,\nn \\ [2mm]
\del_rS_{gen} \sim
-\left( {\phi_r\over 4G}\,{d_i\,t\over r^{d_i+1}} + {c\over 12} {d_i-1\over r}
\right) = 0 ,&&
\del_tS_{gen} \sim
{\phi_r\over 4G}\,{1\over r^{d_i}} + {c\over 12}\,{d_i-1\over d_i\,t} = 0\ .\ \
\eea
Since $c>0$ and $d_i>1$, both contributions in both derivative
expressions appear with the same sign. Note also that in this entire
discussion, as described after (\ref{lr*t*A-2}), we are on one side
(the past) of the singularity at $t=0$, so the range of the time
variable is\ $t\equiv |t|\geq 0$.\ Then it is clear that the only
solution $(t_*,r_*)$ to extremization is
\be\label{AdSKas-t*r*}
r\equiv r_*\ra\infty\ ,\qquad t\equiv t_*\ra\infty\ ;
\qquad t_* \lesssim r_*\ .
\ee
This condition is somewhat analogous to (\ref{t*r*-A=0}).
Since we have assumed $t=t_0$, this is consistent only if
$t_0\ra\infty$ also, \ie\ the observer lies far from the singularity
in the semiclassical spacetime region.  It is useful to note that the
first expression $\del_rS_{gen}$ can be regarded as a spatial
minimization on a fixed Cauchy $t$-slice, which is solved by
$r_*\ra\infty$ for any $t$-value (not growing faster than $r$):
we see that $\del_r^2S_{gen}>0^+$. Then the second expression
$\del_tS_{gen}$, which is a maximization, forces $t_*\ra\infty$\,:
we have\ $\del_t^2S_{gen}\sim -{c\over t^2}\ra 0^+$. It is also
interesting to note from (\ref{SgenAdSKas-t=t0}) that
\be\label{delrtSgen-reln}
\del_rS_{gen} \sim -{d_it\over r} \del_tS_{gen}\ .
\ee
This is a nontrivial relation, which is true only because the 
$t$- and $r$-exponents in $\phi,\, e^f$ are related in a nontrivial
manner. This also explains the last condition on $t_*, r_*$ in 
(\ref{AdSKas-t*r*}): if this were violated, then the extremization
could be violated particularly in $\del_rS_{gen}$. Roughly this is
consistent with preserving the spacelike condition in some sense,
in particular (\ref{Delta^2>0}).
We will expand further on the relation (\ref{delrtSgen-reln}) when
we discuss more general cosmologies later.

In the form (\ref{SgenAdSKas-t=t0}) for $S_{gen}$, it is clear that there
is no Bekenstein bound violating region since the $S_{bulk}$ term is
always subleading to the classical area term as long as $\phi$ is not
small. Thus the spatial extremization in $r$ is expected to lead to
$\phi\ra 0$ or $r_*\ra\infty$.
However it is instructive to note that the $t$-extremization can be
written as
\be
{\del_t\phi\over 4G} + {c\over 6}\,{d_i-1\over d_i} {\del_t\phi\over\phi} = 0\ .
\ee
The fact that $\phi\sim {t\over r^{d_i}}$ with the Big-Crunch
singularity at $t=0$ and noting that both terms here contribute with
the same sign implies that the only solution to extremization is at
$r_*\ra\infty$ from the first term and $t_*\ra\infty$ from the second.
These automatically solve the $r$-extremization as well, as long
as $t_*\lesssim r_*$\,. Another way to see this condition arising
is to note that $S_{bulk}$ positivity implies\ $\phi$ is not too small.
To see this in more detail, let us reinstate lengthscales back: these
are the $AdS$ scale $R$ and the Kasner scale $t_K$ implicit in the
$AdS_D$ Kasner spacetime (\ref{AdSDK-2d}), although $t_K$ does not
play a crucial role. This gives
\be
S_{bulk}\ \sim\ c \log \left( {r^2\over\epsilon_1\epsilon_2}\,
{(t/t_K)^{{d_i-1\over d_i}}\,R^{d_i+1}\over r^{d_i+1}}\,  \right)
\ \sim\
c \log \left(\phi^{{d_i-1\over d_i}}\,{R^2\over\epsilon_1\epsilon_2}\right)
\ee
Thus $S_{bulk}$ becomes negative as $\phi\ra 0$, strictly when $\phi$
becomes smaller than ${R^2\over\epsilon_1\epsilon_2}$\,.  From the higher
dimensional point of view, the area of the transverse space must
become sufficiently small in units of the $AdS$ scale and the
ultraviolet cutoff scales. It is simplest to interpret this as a
breakdown of these expressions in the near singularity region.

It is also interesting to compare this analysis of the generalized
entropy with the classical area term extremization: we have
\be\label{class-Ext-AdSK}
\del_r\phi = -{t\over r^{d_i+1}} = 0\ ,\qquad \del_t\phi = {1\over r^{d_i}} = 0
\ee
both of which are solved by $r_*\ra\infty$, as long as $t_*\lesssim r_*$.
This classical area term does not force the extremal surface to be
driven to $t_*\ra\infty$: in particular we see that $t_*=0$ also
appears formally consistent with these classical extremization
equations. However this entire formulation is unreliable in the
vicinity of $t=0$. The situation is perhaps best appreciated in light
of the higher dimensional RT/HRT analysis earlier: from (\ref{trngpt-tr-2}),
(\ref{lr*t*A-2}), we see that
\be
\phi_*={t_*\over r_*^{d_i}}\sim {1\over l^{d_i}}\ra 0
\ee
in the IR limit $l\ra\infty$\ (see the discussion towards the end of sec.3)
which is the regime probed by the quantum extremal surface in this
2-dimensional analysis here. From the higher dimensional perspective,
the surface dips deep into the bulk so $r_*$ increases: but in the
semiclassical regime as we saw analysing (\ref{del^2rt-2}), the dip in
time is mild, with $t_*\gtrsim t_0$.

Looking more closely, we see that what necessitates driving the
quantum extremal surface to $t_*\ra\infty$ is the ${1\over t}$ term in
(\ref{ExtSgen-r0t0}), (\ref{SgenAdSKas-t=t0}).  This stems from the
power-law Big-Crunch $e^f\sim t^\#$ factor in (\ref{Sgen0}) which
gives a $\log t$ term. If there is no complete Big-Crunch (for
instance as in a bouncing cosmology\footnote{For instance a warp
  factor $e^f\sim (t^2+\delta)^{{d_i-1\over 2d_i}}$ exhibits a bounce
  at $t=0$ without crunching to zero entirely, $\delta$ being a
  regulator. However usually bounces of this kind require violating
  energy conditions or other nonstandard physics. In any case with
  this form of $e^f$ it appears that $t=0$ is also a solution to the
  extremization: it would seem that this is very unreliable since
  we expect severe quantum gravity effects here.}), such a
term would perhaps not arise, allowing finite $t_*$ values.

So far we have been discussing this taking the quantum extremal
surface to be maximally spacelike separated from the observer so
$t=t_0$. This is consistent, as we have seen, with the observer
located far from the singularity since the QES is located far from
the singularity. Now going back to the more general case
(\ref{ExtSgen-r0t0}), it is instructive to note the following. The
first equation in (\ref{ExtSgen-r0t0}) is satisfied for
$r=R_c\sim\infty$, regulating $r_*\ra\infty$ to $r_*\sim R_c$. Then
the second equation can be approximated as
\be
{t-t_0\over R_c^2} \sim {d_i-1\over 2d_i}\,{1\over t}\quad\Rightarrow\quad
t_* \sim {t_0 + \sqrt{t_0^2+4AR_c^2}\over 2}
\ee
which shows $t_*\sim R_c$ as the QES location in (\ref{ExtSgen-r0t0})
regulated from infinity to $R_c\gg 1$. This arises entirely from the
$c$-dependent quantum (bulk) entanglement term.

Now consider the case where the observer is very close to the
singularity, \ie\ $t_0\sim 0$: this is a bad approximation and we
expect a breakdown but it is instructive to analyse this.
Then (\ref{ExtSgen-r0t0}) with $t_0\sim 0$ become
\be
{c\over 6}\,{r\over r^2-t^2} \sim {\phi_r\over 4G}\,{d_i\,t\over r^{d_i+1}}
+ {c\over 12}\,{d_i+1\over r}\,,\qquad
{c\over 6}\,{t\over r^2-t^2} \sim
{\phi_r\over 4G}\,{1\over r^{d_i}} + {c\over 12}\,{d_i-1\over d_i\,t}\ .
\ee
One might wonder if the QES lies in the vicinity of the singularity also,
\ie\ $t=\delta\sim 0$.
The first equation is satisfied if $r_*\ra\infty$ but the second is not
satisfied
due to the last term which diverges as $\delta\ra 0$. This last term
suggests that in fact the QES perhaps lies at $t_*\ra\infty$ again.\
Rather than attempting to look for exact solutions, we will look for
a scaling solution: this will point to the QES again lying far from $t=0$.
Since $\Delta^2=r^2-t^2>0$, consider a trajectory $r=\lambda t$ as an
ansatz for identifying if the QES lies at infinity or at zero (near
the singularity): this is consistent since for both infinity and zero,
the scaling ansatz might be expected to confirm or rule out if the QES
lies at zero, \ie\ in the vicinity of the singularity. This gives
\be
{c\over 12}\,\left({2\lambda\over \lambda^2-1} - {d_i+1\over\lambda}\right)
{1\over t} =  {\phi_r\over 4G\lambda^{d_i+1}}\,{d_i\over t^{d_i}}\,,\qquad
{c\over 12}\,\left({2\over \lambda^2-1} - {d_i-1\over d_i}\right)
{1\over t} =  {\phi_r\over 4G\lambda^{d_i}}\,{1\over t^{d_i}}\ .
\ee
In general, the coefficients of the ${1\over t}$ and ${1\over t^{d_i}}$
terms are different\footnote{For instance, as we ``dial'' $\lambda$
  from $\lambda\gg 1$ to $\lambda=1+\varepsilon$, we have
\be
{c\over 12}\,{1-d_i\over\lambda}\,{1\over t} \sim
{\phi_r\over 4G\lambda^{d_i+1}}\,{d_i\over t^{d_i}}\,,\ \ 
{c\over 12}\,{1-d_i\over d_i}\,{1\over t} \sim
{\phi_r\over 4G\lambda^{d_i}}\,{1\over t^{d_i}}\,;\qquad 
{c\over 12}\,{1\over\varepsilon}\,{1\over t} \sim
{\phi_r\over 4G}\,{d_i\over t^{d_i}}\,,\ \
{c\over 12}\,{1\over\varepsilon}\,{1\over t}
\sim {\phi_r\over 4G}\,{1\over t^{d_i}}\,,\nn
\ee
vindicating the $t_*\ra\infty$ solution.}:
so each term must vanish independently: thus the only solution is
$t_*\ra\infty$ and thereby $r_*\ra\infty$.

Thus this formulation of the generalized entropy appears to
self-consistently exclude the near singularity region. In a sense the
fact that the quantum extremal surfaces are driven far from the
singular region is reassuring with regard to the Page curve findings
\cite{Penington:2019npb,Almheiri:2019psf,Almheiri:2019hni} which
appear to not require any information from the near singularity
region.

\subsection{More general 2-dim cosmologies}

We now make a few comments on the generalized entropy and quantum
extremal surfaces in more general 2-dim cosmologies with the general
scaling form (\ref{phie^fPsi-ansatz}), defined by various $t$- and
$r$-exponents used in \cite{Bhattacharya:2020qil}.  We will also
assume for simplicity that the QES is maximally spacelike separated
from the observer, thus setting $t=t_0$ in the general expression
(\ref{Sgen0}). Finally since all these cosmologies have a boundary at
$r=0$ we will restrict attention to such boundary observers. This
gives
\be\label{Sgen-genCos}
S_{gen} = {\phi\over 4G} + {c\over 12} \log \left(r^2\, e^f|_{(t,r)} \right)
= {t r^m\over 4G} + {c\over 12} \log \left(t^a r^{b+2} \right)
\ee
retaining only terms relevant for the extremization as before. From
the various examples in Sec.~\ref{sec:rev2dCos}, we know that $m, b<0$:
this is in accord with the transverse space, \ie\ dilaton, expanding
towards the boundary $r\sim 0$. Further we have also used the
universality of the time exponent of the dilaton.
Firstly, this can be recast in the schematic form (\ref{SgenAdSKas-t=t0})
only if the argument of the logarithm is related to the dilaton exponents
appropriately: this gives
\be
S_{gen}\equiv {\phi\over 4G} + {c\,a\over 12} \log \phi
\qquad\Leftrightarrow\qquad
a = {2+b\over m}\ .
\ee
It is interesting to note that this relates the $t$- and $r$-exponents
which were otherwise independent in general in \cite{Bhattacharya:2020qil}.
Extremization of (\ref{Sgen-genCos}) gives
\be\label{Sgen-gndSt}
\del_rS_{gen} = {mt r^{m-1}\over 4G} + {c\over 12}\, {2+b\over r} = 0\ ,
\qquad
\del_tS_{gen} = {r^m\over 4G} + {c\over 12}\, {a\over t} = 0\ ,
\ee
\be\label{Sgen-gndSt-reln}
\Rightarrow\qquad\qquad
\del_rS_{gen} \sim {mt\over r} \del_tS_{gen} \qquad\Leftrightarrow\qquad
a = {2+b\over m}\ .\qquad
\ee
From the time extremization $\del_tS_{gen}=0$ and noting $m<0$ we see
that the solution to the QES location is $r_*\ra\infty,\ t_*\ra\infty$,
with $t_*\lesssim r_*$.
This structure is similar to (\ref{AdSKas-t*r*}) for the 2-dim
cosmology obtained from the $AdS$ Kasner reduction that we discussed
earlier.
Now looking at just the classical area term, with a scaling form
$\phi=t^kr^m$, extremizing gives
\be
\del_r\phi = mt^kr^{m-1}=0\ ,\qquad  \del_t\phi = kt^{k-1}r^m = 0\ .
\ee
Since the dilaton grows towards the boundary $r\ra 0$, we must have
$m<0$.\ Then for $k>0$, these equations are similar to
(\ref{class-Ext-AdSK}), with both satisfied if $r_*\ra\infty$ as long
as $t_*\lesssim r_*$. With the universality (\ref{univSing}), taking
$k=1$ so $\phi=tr^m$, a general bulk matter entropy $S_{bulk}$ gives
\be
S_{gen} = {t\,r^m\over 4G_2} + S_b\,;\qquad
\del_rS_{gen} = {m\,t\,r^{m-1}\over 4G_2} + \del_rS_b = 0\ ,\quad
\del_tS_{gen} = {r^m\over 4G_2} + \del_tS_b = 0\ .
\ee
For $S_b$ being the ground state entanglement, we obtain
(\ref{Sgen-gndSt}), (\ref{Sgen-gndSt-reln}) above, for the scaling
form. More generally, a relation of the form (\ref{Sgen-gndSt-reln})
arises for $S_{gen}$ if $S_{bulk}$ satisfies\
$\del_rS_b = {m\,t\over r}\,\del_tS_b$.
For instance an extensive bulk entropy $S_{bulk}$ of the form below gives
\bea
&& S_{bulk} = \Lambda\,r\,e^{f/2}|_{(t,r)} = \Lambda\,t^{a/2}\,r^{(2+b)/2}
\qquad\ra\qquad \nn\\
&& \del_rS_{gen} = {m\,t\,r^{m-1}\over 4G_2} + \Lambda\,t^{a/2}\,r^{b/2} = 0\ ,
\quad
\del_tS_{gen} = {r^m\over 4G_2} + \Lambda\,t^{(a-2)/2}\,r^{(2+b)/2} = 0\ ,
\eea
thus satisfying the relation (\ref{Sgen-gndSt-reln}) if
$a={2+b\over m}$\,. In particular this is true for the $AdS$ Kasner
reduction earlier. However this generalized entropy can be seen to
vanish at the location $t=0$ of the singularity: it is unclear if this
is reasonable. It would be interesting to explore good models for near
singularity physics and the resulting quantum extremal surfaces.


\section{Discussion}

We have studied aspects of entanglement and extremal surfaces in
various families of spacetimes exhibiting cosmological, Big-Crunch,
singularities, in particular the isotropic $AdS$ Kasner spacetime.
The classical RT/HRT extremal surface dips into the bulk radial and
time directions, with turning points (\ref{trngpt-tr-2}),
(\ref{lr*t*A-2}), satisfying\ $l\sim r_*$ and $\phi_*=A={t_*\over
  r_*^{d_i}}$, for $AdS_{d_i+2}$ Kasner.  By analysing the time
extremization equation in the reliable semiclassical region far from
the singularity via (\ref{del^2rt-2}), we have seen explicitly that
the surface lies mostly on a constant time slice and bends in the
direction away from the singularity at $t=0$. At the turning point,
the surface exhibits time-maximization. As we have seen, the IR limit
where $A\ra 0$ continues to exhibit such behaviour, but also shows
indications of the analysis breaking down. In the 2-dim cosmologies
obtained by dimensional reduction of these and other singularities
\cite{Bhattacharya:2020qil}, we have studied quantum extremal
surfaces. The generalized entropy (\ref{Sgen0}) comprises the
classical area (dilaton) term and a bulk matter entropy obtained by
using the formulation of \cite{Penington:2019npb,Almheiri:2019psf}
with the effects of the curved space incorporated via the conformal
transformation, taking the matter in the ground state in flat space
(in the region far from the singularity at $t=0$). The resulting
extremization shows the quantum extremal surfaces exhibit a maximin
structure: they are always driven to the semiclassical region far from
the singularity\ (\eg\ (\ref{SgenAdSKas}), (\ref{SgenAdSKas-t=t0}) for
the isotropic $AdS$ Kasner reduction).  Technically this follows from
the Crunching term in the warp factor. We do not find islands in this
analysis: this appears consistent with previous investigations on
closed universes, and can be interpreted in terms of the Bekenstein
bound not being violated \cite{Hartman:2020khs}, so that there is no
competition for the area (dilaton) term. It would be interesting to
consider extra regions elsewhere (\eg\ flat space regions beyond the
boundary $r=0$ or in the far past) that are entangled with these
universes: these may exhibit islands.

In the discussion of quantum extremal surfaces, we have used 2-dim CFT
techniques: these are technically reasonable in the 2-dim theories
assuming that the bulk matter is described by a CFT. These 2-dim
backgrounds are consistent intrinsically as solutions to the 2-dim
dilaton-gravity-scalar theories (\ref{actionXPsiU}), so the
formulation of generalized entropy here is consistent in the
semiclassical regime. These would seem to faithfully capture
qualitative features of quantum extremal surfaces in the higher
dimensional cosmologies that give rise to the 2-dim backgrounds upon
dimensional reduction, at least considering that the surfaces are
driven to the semiclassical region. In some ``effective holography''
sense, the 2-dim backgrounds faithfully reflect the higher dimensional
description (see \cite{Narayan:2020pyj} for more discussions in this
regard). However it would seem that this would break down had the
vicinity of the singularity entered: happily the quantum extremal
surfaces avoid this.

In a sense these are reminiscent of similar features noted in the
study of the Hartman-Maldacena extremal surfaces
\cite{Hartman:2013qma} in the $AdS$ black hole where the extremal
surface approached a limiting surface some distance from the
singularity in the interior\ (similar limiting surfaces were found in
\cite{Narayan:2020nsc} in de Sitter: it would be interesting to
understand quantum extremal surfaces in that context); similar
observations were noted also in the $AdS$ Kasner soliton
\cite{Engelhardt:2013jda}, and other cases. We offer some comments and
speculations on these results. Perhaps the simplest understanding
hinges on the fact that the near singularity region is necessarily a
place where quantum gravity effects are severe and the formulation of
quantum extremal surfaces, such as it is, is simply not adequate. A
more detailed model of the near singularity region incorporating
perhaps ``stringy entanglement'' may be necessary. Our studies here
seem consistent with the recent excitement on black holes and the Page
curve: all the action there remained well separated from the near
singularity region in the deep interior of the black hole (this
singularity is anisotropic Kasner).  In a sense the study here is
reassuring since the quantum extremal surfaces self-consistently avoid
the near singularity region rife with quantum gravity effects,
remaining in semiclassical regimes far away. Turning this around, one
might speculate if such Big-Crunch singularities are perhaps
disallowed in string theory and holography, based on the (naive)
diagnostic that entanglement via extremal surfaces is strictly
incapable of probing the vicinity of such singularities.  It is worth
noting that the models we have studied pertain to $AdS$-cosmologies
and related backgrounds\ (the universality (\ref{univSing}) found in
\cite{Bhattacharya:2020qil} suggests that the singularity nature of
all such backgrounds is similar).  These have a timelike boundary at
$r=0$ with a holographic dual field theory interpretation: the
extremal surfaces we have discussed thus encode entanglement
observables in the dual field theories.  It may be interesting to
understand if more general cosmologies without such holographic
restrictions exhibit similar features. It is also important to note
that other holographic cosmologies may not exhibit this sort of
repulsive behaviour: an example is anisotropic $AdS$ Kasner with some
directions expanding and some Crunching, but this lies outside the
reduction to the 2-dim space we have employed.

Relatedly it may also be worth noting that perhaps assuming that bulk
matter is in its ground state far from the singularity in our models
is a nontrivial assumption\footnote{This is in certain cases
  consistent with a Euclidean continuation which might lead to a
  natural initial state\ (although generic time-dependent spacetimes
  become complex).  For instance de Sitter space
  $ds^2={R_{dS}^2\over\tau^2}(-d\tau^2+dx_i^2)$ under the analytic
  continuation $\tau\ra ir ,\ R_{dS}\ra iR_{AdS}$ becomes Euclidean
  $AdS$ with corresponding continuations for fields and their boundary
  conditions; see also \eg\ \cite{Cooper:2018cmb,
    VanRaamsdonk:2020tlr} in other contexts.  Then we see that
  (\ref{AdSDK-2d}) naively admits a similar Euclidean continuation
  $t\ra i\tau ,\ t_K\ra i\tau_K$, if we also analytically continue the
  Kasner scale $t_K$: equivalently, on the $t=t_K$ time slice
  sufficiently far from the singularity, we could consider gluing
  \eg\ a flat region, in part along the lines of \cite{Chen:2020tes},
  which appears consistent with taking matter in the ground state. We
  hope to study this in greater detail in future work towards
  understanding initial conditions for spacetimes developing such
  Big-Crunch singularities.}: generic initial Cauchy data might be
expected to lead to black hole formation rather than a Big-Crunch
singularity (further discussions appear in
\cite{Bhattacharya:2020qil}).  From this point of view, it is perhaps
not surprising that the generalized entropy incorporating ground state
matter entanglement leads to quantum extremal surfaces that avoid the
singularity. Perhaps a better model might incorporate more nontrivial
initial conditions for the bulk matter and the associated entropy
would then naturally lead to a Big-Crunch singularity that is
accessible via entanglement.  It would be interesting to explore these
issues further.

\vspace{5mm}

{\footnotesize \noindent {\bf Acknowledgements:}\ \ It is a pleasure
  to thank Tom Hartman for a very useful early discussion, and Tom
  Hartman, Arnab Kundu and Raghu Mahajan for comments on a draft. We
  thank Ritabrata Bhattacharya for collaboration in the initial stages
  of this work, as well as Alok Laddha for discussions. This work is
  partially supported by a grant to CMI from the Infosys Foundation.
}

\appendix

\section{Further details on $t(r)$ for classical extremal surfaces}

We shall describe some relevant details of the behaviour of
the extremal surface in the higher dimensional case.\ 
In the $AdS_{5}$ case, the $t(r)$ function, as a power series solution
to (\ref{del^2rt-2}), is
{\footnotesize
\bea    \label{AdS5-t(r)-A}
&& t(r) =t_{0} +\frac{1}{12t_{0}}r^{2}-\frac{1}{432t_{0}^{3}}r^{4} +\frac{1}{7776t_{0}^{5}}r^{6} + \frac{(-17+11664A^{2}t_{0}^{4})}{1866240t_{0}^{7}}r^{8}
+ \frac{(247-400464A^{2}t_{0}^{4})}{335923200t_{0}^{9}}r^{10} + \qquad \nn\\
&&\qquad \frac{(-1819+5110128A^{2}t_{0}^{4})}{28217548800t_{0}^{11}}r^{12} + \frac{(21277-90004284A^{2}t_{0}^{4} + 5555329920A^{4}t_{0}^{8})}{3555411148800t_{0}^{13}}r^{14} + \nn\\
&&\qquad \frac{(-373318170624 A^4t_0^8+2088752184 A^2t_0^4-355981)}{614375046512640 t_0^{15}}\,r^{16} + \nn\\
&&\qquad \frac{(257320977209856 A^4t_0^8-740910019032 A^2t_0^4+96110087)}{1658812625584128000 t_0^{17}}\,r^{18} + \\
&& \frac{(707587260600729600 A^6t_0^{12}-35768655789931008 A^4t_0^8+63403487354664 A^2 t_0^4-6506954915)}{1094816332885524480000 t_0^{19}}\,r^{20}
+ \ldots \nn
\eea
}
One can compute higher order coefficients iteratively using Mathematica.

We shall now describe how the different terms in the above series behave
under the scaling argument described after eq.\eqref{AdS5-t(r)}.\ 
As $At_{K}$ is $t_{*}/r_{*}^{3}$ and since $t_{*}\sim t_{0} \Rightarrow A\sim t_{0}/r_{*}^{3}$. Using this scaling of $A$, we can see how different terms in the above equation scale.\\
For instance, the $A$ dependent term at $r^{8}$, after multiplying and dividing by $r_{*}^{2}$ and pluggin in the above scaling for $A$, is $\frac{A^{2}}{t_{0}^{3}}r^{8}\sim (r_{*}^{2}/t_{0}) (r/r_{*})^{8}$. Similarly the $A^{4}$ term at $r^{14}$ is $\frac{A^{4}}{t_{0}^{5}}r^{14} \sim (r_{*}^{2}/t_{0}) (r/r_{*})^{14}$. Whereas the $A^{2}$ dependent terms at $r^{10}, r^{12}$ and $r^{14}$ go as $\frac{A^{2}}{t_{0}^{5}}r^{10}\sim (r_{*}^{4}/t_{0}^{3})(r/r_{*})^{10}$, $\frac{A^{2}}{t_{0}^{7}}r^{12}\sim (r_{*}^{6}/t_{0}^{5})(r/r_{*})^{12}$ and $\frac{A^{2}}{t_{0}^{9}}r^{14}\sim (r_{*}^{8}/t_{0}^{7}) (r/r_{*})^{14}$, respectively.\\
There are a few points which are noteworthy about these $A$ dependent terms which we describe below.\\
First and most importantly, this scaling argument shows that all the $A$ dependent terms are also suppressed as compared to the leading term in the series which is the first term.\\
The second point is that the terms at a particular order of $r$ can be compared with each other and the $A$ dependent terms are the ones which dominate within these terms, e.g, there are two terms at $r^{8}$ which are $\frac{r^{8}}{t_{0}^{7}} \sim (r_{*}^{8}/t_{0}^{7})(r/r_{*})^{8}$ and $\frac{A^{2}}{t_{0}^{3}}r^{8}\sim (r_{*}^{2}/t_{0}) (r/r_{*})^{8}$. Since the former term has a higher power of $t_{0}$ in the denominator and since, $t_{0}\gg 1$, the latter term is dominant.\\
The final point is that at each order of $2d_{i}$, a $A^{2k}$ term appears where $k=1,2...$. This can be seen in the above case of $AdS_{5}$ where $A^{2}$, $A^{4}$, $A^{6}$ etc. appear at $r^{8},r^{14}, r^{20}$, respectively.
Keeping the dominant terms amongst each power of $r$ gives
{\footnotesize
\begin{multline}
t(r) =t_{0} +\frac{1}{12t_{0}}r^{2}-\frac{1}{432t_{0}^{3}}r^{4} +\frac{1}{7776t_{0}^{5}}r^{6} + \frac{A^{2}}{160t_{0}^{3}}r^{8}
- \frac{103A^{2}}{160\cdot 540t_{0}^{5}}r^{10}
+ \frac{3943A^{2}}{160\cdot 540\cdot 252t_{0}^{7}}r^{12}\\
+ \frac{A^{4}}{160\cdot 4t_{0}^{5}}r^{14}
- \frac{7A^{4}}{160 \cdot 72t_0^{7}}\,r^{16}
+ \frac{15011 A^{4}}{160\cdot 540 \cdot 1120t_0^{9}}\,r^{18}
+ \frac{91A^6}{160\cdot 880\,t_0^{7}}\,r^{20}\\
+ \frac{8453 A^6}{22302720\,t_0^9}\,r^{22}
+ \frac{493338049 A^6}{3653185536000\,t_0^{11}}\,r^{24}
+ {19 A^8\over 56320 t_0^9}\,r^{26} + \ldots \qquad
\end{multline}
}
However at $r_{*}$ various terms become subleading using
$A={t_*\over r_*^3}$\,: \eg\ we see that the $r_*^{10}$ and $r_*^{12}$
terms are suppressed by powers of $t_0$ and thus subleading compared
with the $r_*^8$ term, all scaling as $A^2$. Now the $r_*^{14}$
term has scaling\ ${t_*^4\over r_*^{12}}\,{r_*^{14}\over t_0^5}\sim
{r_*^2\over t_0}$ which is the same as the $r^8$ term: the higher
terms \eg\ $r_*^{20}$ also has the same scaling, and so on. Thus the
above series is further approximated as
\begin{equation}\label{t(r)-sub}
t(r) = t_0 + \frac{r_{*}^{2}}{t_{0}}\bigg( {1\over 12 }\, + {1\over 160 }\,
+ {1\over 160\cdot 4}\, + {91 \over 160 \cdot 880 }\, +
{19 \over 160\cdot 352}\, + \ldots\bigg)
\end{equation}
Thus we recover (\ref{t*>t0}), \ie\ $t_{*} > t_{0}$ when $r_*\lesssim t_0$.

All the arguments above hold for $AdS_{4}$ and $AdS_{7}$ Kasner as well.
For $AdS_{4}$, we obtain 
{\footnotesize
\begin{multline}
t(r) = t_{0} +\frac{1}{12t_{0}}r^{2} -\frac{1}{480t_{0}^{3}}r^{4} + \frac{(13 +1920A^{2}t_{0}^{2})}{120960t_{0}^{5}}r^{6} \\ -\frac{(125 +44928A^{2}t_{0}^{2})}{17418240t_{0}^{7}}r^{8} + \frac{(10543 +6641280A^{2}t_{0}^{2} + 82944000A^{4}t_{0}^{4})}{19160064000t_{0}^{9}}r^{10} + \ldots
\end{multline} }
and for $AdS_{7}$
{\footnotesize
\begin{multline}
t(r) = t_{0} + \frac{1}{15t_{0}}r^{2} + \frac{1}{600t_{0}^{3}}r^{4}+\frac{11}{135000t_{0}^{3}}r^{6}- \frac{1}{200000t_{0}^{7}}r^{8} + \frac{491}{1417500000t_{0}^{9}}r^{10} \\+ \frac{(-52891+3543750000A^{2}t_{0}^{8})}{128595600000000t_{0}^{13}}r^{14}+ \ldots
\end{multline} }

\section{Some details on calculating $S_{bulk}$}

Since any 2-dim metric is conformally flat, we have\ $ds^2=e^f(-dt^2+dr^2)$.
If we now assume that the bulk matter can be modelled by a 2-dim CFT,
we can obtain its entropy as in \cite{Almheiri:2019psf} by modifying
the Calabrese-Cardy result \cite{Calabrese:2004eu,Calabrese:2009qy},
in particular taking the ground state entanglement in flat space and
incorporating the effects of the Weyl transformation $e^f$. The twist
operator 2-point function scales under a conformal transformation as 
\be
\lan \sigma(x_1)\,\sigma(x_2)\ran_{e^fg} = e^{-\Delta_n\,f}\vert_{x_1}\,
e^{-\Delta_n\,f}\vert_{x_2}\, \lan \sigma(x_1)\,\sigma(x_2)\ran_{g}\ ,\qquad
\Delta_n={c\over 12} {n^2-1\over n}\ .
\ee
Since the partition function in the presence of twist operators scales
as the twist operator 2-point function, the entanglement entropy becomes
\be
S^{12}_{e^fg} = -\lim_{n\ra 1}\,\del_n \lan \sigma(x_1)\,\sigma(x_2)\ran_{e^fg}
= S^{12}_g - {c\over 6} \log e^f\vert_{x_1} - {c\over 6} \log e^f\vert_{x_2}
\ee
For the interval taken with one endpoint at the boundary, we essentially
have a single twist operator and its 1-point function, using the
boundary CFT prescription: this essentially
restricts to just the single boundary in the bulk, say $x_1$, and
effectively $c\ra {c\over 2}$ above. We consider the interval
$(r,r_c)\sim (r,0)$ with endpoints being the extremal surface at $r$
and the boundary $r_c\sim 0$, the whole space being the half-line
$(\infty,0)$. We construct the replica space (with
$w\equiv \tau+ix$) for this situation by gluing the $n$-sheets along
the cut defined by the interval with twist operators at the single
endpoint. To analyse the replica theory, we first map the half-line
to a disc $|z|\leq 1$ via the uniformization map\
$z=({w-il\over w+il})^{1/n}$\ which maps the boundary $w=0$ to $z=1$
and the endpoint at $x=l$ (so $w=ix=il$) to $z=0$. There is a single
twist operator at the endpoint $x=l$ \ (unlike two for a single interval
in the full line). Taking the $z$-plane to be the $SL(2)$ vacuum so
$\lan T(z)\ran=0$, we find the expectation value of the stress tensor
$\lan T(w)\ran_{R_n}$ via the Schwarzian for the $z(w)$ map above:
this gives\  $\lan T(w)\ran_{R_n} = {c\over 24} (1-{1\over n^2})
{(2l)^2\over (w-il)^2(w+il)^2}$. Using BCFT Ward identities etc, this
is equivalent to\
${\lan T(w) \sigma_n(il)\ran\over \lan \sigma_n(il)\ran}$
with $\lan \sigma_n(il)\ran={1\over (2l)^{c(n-1/n)/12}}$\,.
Then the replica partition function transforms as\
$Tr\rho_A^n\sim \lan \sigma_n(il)\ran$\ and
$S_A=-\lim_{n\ra 1}\del_nTr\rho_A^n = {c\over 6}\log {2l\over\epsilon}$\,.
Roughly this is like half the area (one endpoint rather than two for
an interval in the full line space).\ 
So the expression in (\ref{Sgen0}) is written for the 2-dim space with
the boundary at $r=r_c$ and the QES at $(t,r)$.

Since the conformal factor has nontrivial time-dependence, the
conformal transformation above is nontrivial. In writing
(\ref{Sgen0}), we are making the nontrivial assumption that this
formulation can be applied: this appears reasonable in the
semiclassical region far from the singularity.  However the presence
of the Big-Crunch at $t=0$ as $e^f\ra 0$ is expected to lead to a
breakdown of this formulation, as we have stated in the text.

\end{document}